\documentclass[aps,twocolumn,superscriptaddress,showpacs,amssymb,prl,floatfix,longbibliography]{revtex4-2}
\usepackage{graphicx }
\usepackage[]{hyperref}
\usepackage{dcolumn}
\hypersetup{colorlinks=true,linkcolor=blue,citecolor=blue,urlcolor=blue,pdfpagemode=UseNone}

\newcommand{\slrrtext}  {spin-lattice-relaxation rate}
\newcommand{\slrr}      {$T_1^{-1}$}

\begin{document}

\title{Inhomogeneous Knight shift in vortex cores of superconducting FeSe}

\author{I. Vinograd}
    \affiliation{Department of Physics and Astronomy, University of California, Davis, California
95616, USA}
\author{S. P. Edwards}
    \affiliation{Department of Physics and Astronomy, University of California, Davis, California
95616, USA}
\author{Z.  Wang}
    \affiliation{Department of Physics and Astronomy, University of California, Davis, California
95616, USA}
\author{T. Kissikov}
    \affiliation{Department of Physics and Astronomy, University of California, Davis, California
95616, USA}
\author{J. K. Byland}
    \affiliation{Department of Physics and Astronomy, University of California, Davis, California
95616, USA}
\author{J. R. Badger}
    \affiliation{Department of Chemistry, University of California, Davis, California
95616, USA}
\author{V. Taufour}
    \affiliation{Department of Physics and Astronomy, University of California, Davis, California
95616, USA}
\author{N. J. Curro}
    \affiliation{Department of Physics and Astronomy, University of California, Davis, California
95616, USA}
\date{\today}
\begin{abstract}

We report $^{77}$Se NMR data in the normal and superconducting states of a single crystal of FeSe for several different field orientations.  The Knight shift is suppressed in the superconducting state for in-plane fields, but does not vanish at zero temperature. For fields oriented out of the plane, little or no reduction is observed below $T_c$. These results reflect spin-singlet pairing emerging from a nematic state with large orbital susceptibility and spin-orbit coupling. The spectra and spin-relaxation rate data reveal electronic inhomogeneity that is enhanced in the superconducting state, possibly arising from enhanced density of states in the vortex cores. Despite the spin polarization of these states, there is no evidence for antiferromagnetic fluctuations.

\end{abstract}

\maketitle

The iron-based superconductors have attracted broad interest recently because they can host Majorana modes on the surface, at domain walls, and within vortex cores \cite{Koenig2019, Zhang2018,Kong2020,Wang2020}.  Fe(Se,Te), and Li(Fe,Co)As contain bands with $p_z$ and $d_{xz}/d_{yz}$ character with non-trivial topologies, that give rise to both topological surface states as well as a bulk Dirac point near the $\Gamma$ point in k-space \cite{Wang2015}.   FeSe, although topologically trivial, is a particularly interesting case because the superconducting state emerges from a nematic phase that develops below $T_{nem} = 91$K \cite{Kreisel2020}. Moreover, the Fermi energy, $E_F$, in this system is usually  small, such that this system lies close to the BCS-BEC crossover regime \cite{Hanaguri2019,Shibauchi2020}.  Evidence has emerged that suggests FeSe exhibits a Fulde-Ferell-Larkin-Ovchinnikov (FFLO) phase at high magnetic fields \cite{Kasahara2020}. The possibility of both FFLO and  dispersive Majorana modes underlies the importance of a detailed understanding of the nature of the vortices in these materials.

To probe vortex matter it is important to first understand the underlying superconducting state. The spatial part of the superconducting wavefunction in FeSe is generally assumed to be either $s_{\pm}$ or $d$-wave. The nematic normal state gives rise to twin domains and in-plane anisotropy, and the Fermi surfaces contain different orbital characters in the two domains.  The superconducting gap, $\Delta$, appears to correlate with the orbital content on the Fermi surface \cite{FeSeOrbitalPairing2017}. However, the presence of the domains may mask intrinsic properties about the density of states below $T_c$, and there are conflicting reports about the presence or lack of nodes and anisotropy of the superconducting gap function  \cite{BourgeoisHope2016,Jiao2017,Biswas2018,Hardy2019}.

\begin{figure}
\begin{center}
\includegraphics[width=\linewidth]{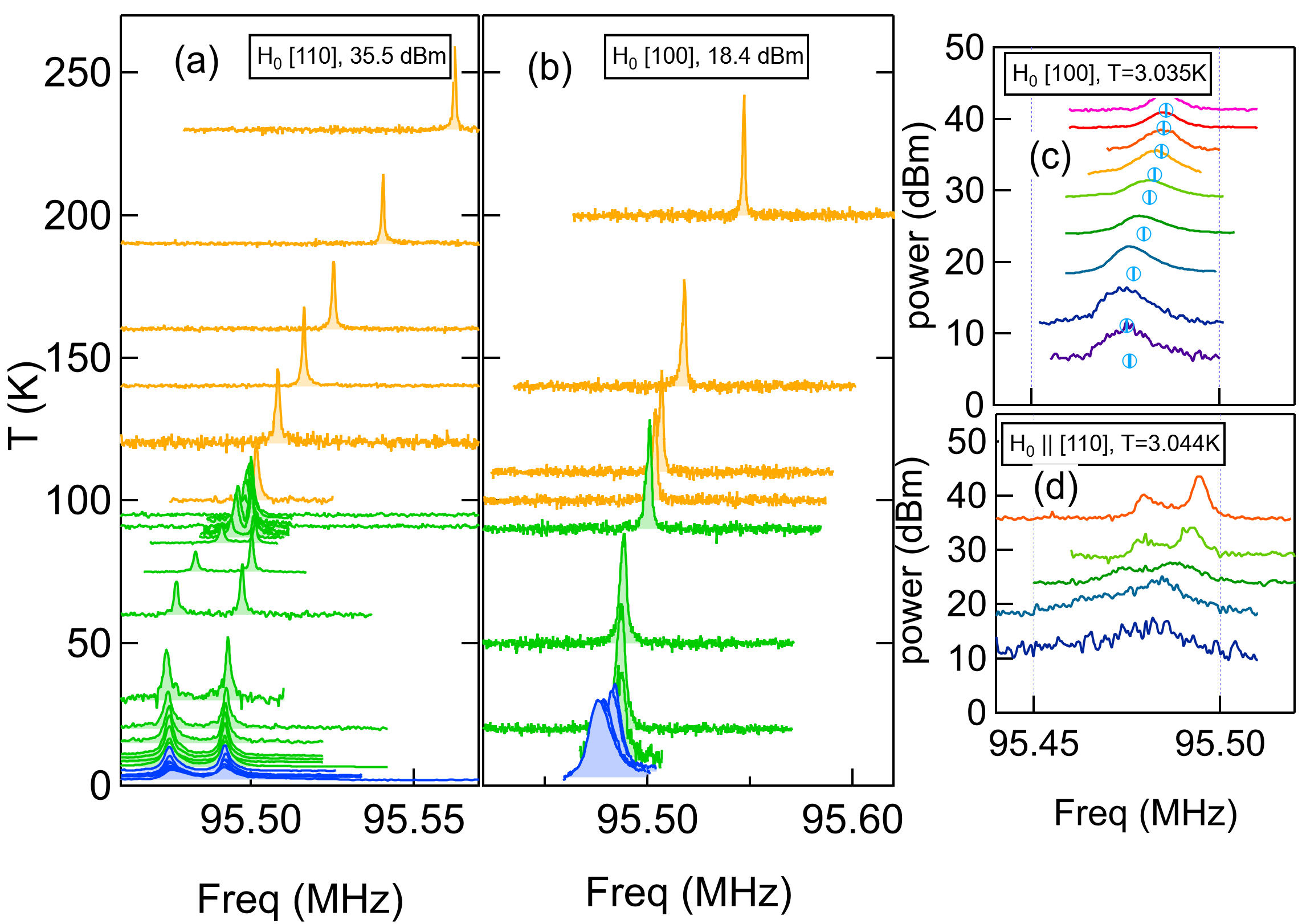}
\caption{(a) $^{77}$Se NMR spectra (normalized) as a function of temperature for $\mathbf{H}_0 \parallel[110]$ (in tetragonal unit cell) at high rf power (35.5 dBm).  Below $T_{nem} = 91$K, the single resonance splits into two separate peaks, corresponding to domains with $\mathbf{H}_0\parallel\mathbf{a}$ (upper peak) and $\mathbf{H}_0\parallel\mathbf{b}$ (lower peak) in the nematic phase. (b) Spectra as a function of temperature for $\mathbf{H}_0 \parallel[110]$ at low rf power (18.4 dBm). (c,d) Spectra in the superconducting state as a function of radiofrequency pulse power for $\mathbf{H}_0\parallel [100]$ and  $\mathbf{H}_0\parallel [110]$, respectively. Blue circles in (c) indicate the first moment of the spectrum.}
\label{fig:waterfall}
\end{center}
\end{figure}

Information about the spin component of the wavefunction can be gleaned from  nuclear magnetic resonance (NMR) Knight shift measurements.  The spin susceptibility of a condensate with singlet pairing vanishes, whereas that with triplet pairing can remain unchanged through $T_c$. Conventional magnetometry cannot discern these changes because the spin component is much smaller than the orbital component, however the Knight shift is usually dominated by the former
and is thus one of the only experimental probes of the spin susceptibility of the condensate
\cite{Hanaguri2019,Shibauchi2020}. To date, Knight shift measurements in the superconducting state have been inconclusive, revealing little or no change below $T_c$ \cite{Baek2015,Wiecki2017,Baek2016,Li2020}.  A recent study reported no change in the Knight shift along the $c$ axis in fields up to 16 T, which have been interpreted as evidence for highly spin-polarized Fermi liquid in the BCS-BEC regime \cite{Molatta2020}.  A lack of suppression of the  Knight shift may be evidence for spin-triplet pairing \cite{IshidaSr2RuO4}, but may also reflect thermal instability of the sample due to eddy-current heating from radiofrequency pulses \cite{Pustogow2019}.   In fact,  spin-orbit coupling can give rise naturally to a spin-triplet component \cite{Vafek13,Vafek2017}.  To fully characterize the symmetry of the condensate, therefore, it is important to understand the full tensor nature of the Knight shift in the superconducting state.

Here we report $^{77}$Se NMR on a high quality single crystal as a function of temperature and field. We find that between 3.6 and 11.7 T, the spin part of the planar Knight shift is reduced by $\sim 10-15\%$ from their normal state values below $T_c$, whereas the out-of-plane component shows no change within the experimental resolution. These results are consistent with spin singlet pairing in the presence of large orbital susceptibility and spin-orbit coupling. Surprisingly, the NMR linewidths broaden inhomogenously by more than a factor of two below $T_c$ for planar fields, but not for $\mathbf{H}_0\parallel \mathbf{c}$.  Accompanying this broadening is a frequency-dependent spin-lattice relaxation rate, \slrr, that reveals electronic inhomogeneity in the superconducting state. This inhomogeneity cannot be explained by the presence of a conventional vortex lattice, but may reflect an enhanced local density of states within the vortex cores.

\begin{figure}[h]
\begin{center}
\includegraphics[width=\linewidth]{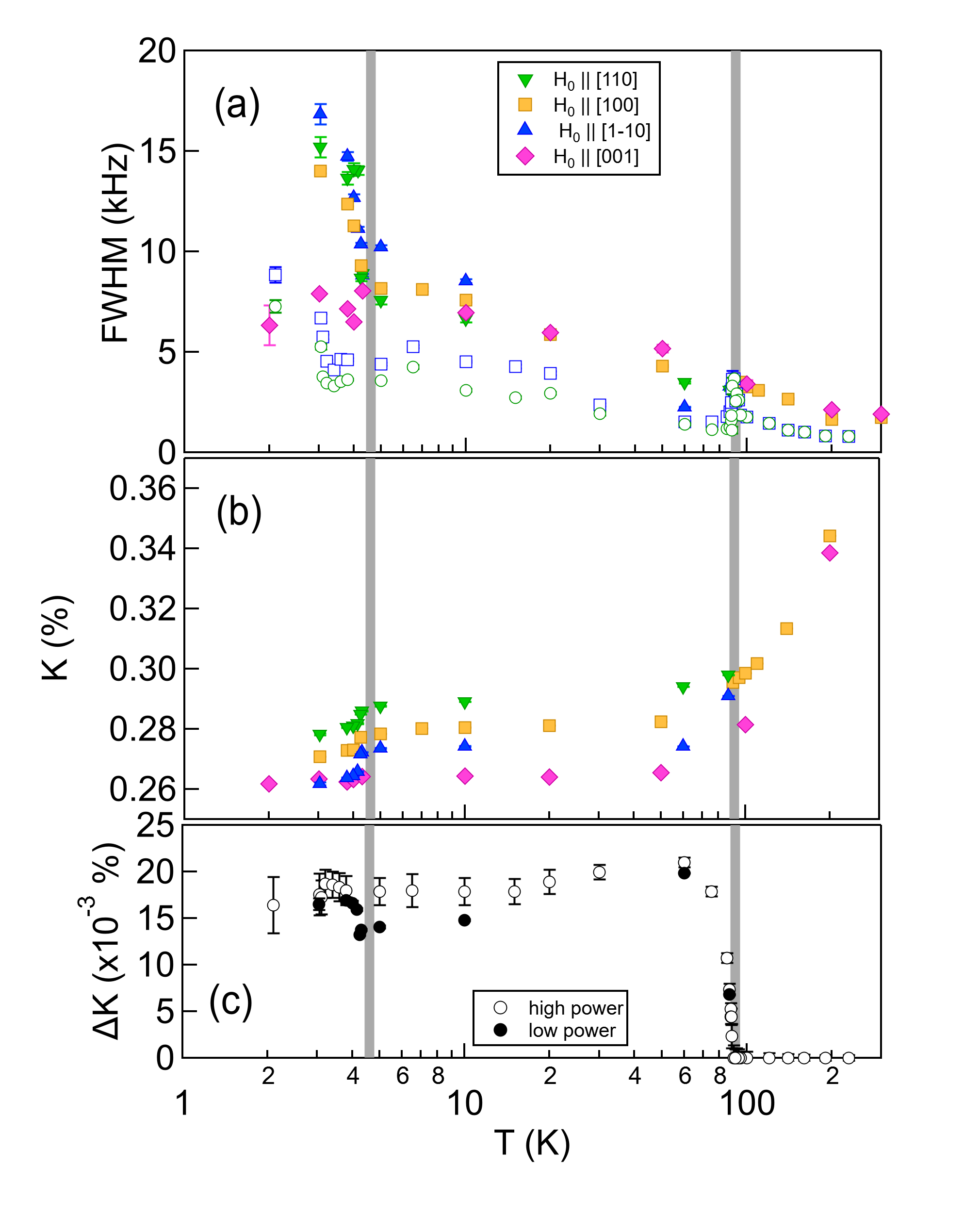}
\caption{Linewidth (a) and Knight shifts (b) of the spectra in Fig. \ref{fig:waterfall} as a function of temperature for the field along $[110]\sim\mathbf{a}$ ($\blacktriangledown$), $[1\bar{1}0]\sim\mathbf{b}$ ($\blacktriangle$), $[100]$ ($\blacksquare$), and $[001]\sim\mathbf{c}$ ($\blacklozenge$). (c) The in-plane anisotropy $\Delta K=K_a - K_b$ as a function of temperature. Solid (open) points were acquired at low (high) rf power, respectively, as discussed in the text. }
\label{fig:width}
\end{center}
\end{figure}

Single crystals of FeSe were grown by vapor transport with a tilted two-temperature zone tube furnace~\cite{Boehmer2016}. Several samples were characterized with magnetic susceptibility and resistivity measurements, with the best samples having $T_c=8.9$\,K, and RRR defined as the resistance ratio between 250 and 10\,K around 19, similar to reported high-quality samples~\cite{Boehmer2016}. A large crystal of dimensions 2.4 mm $\times$1.4 mm $\times$ 0.2 mm was selected and mounted in a custom-built NMR probe equipped with a dual-axis goniometer. The majority of the experiments were conducted within a variable-temperature cryostat in a high-homogeneity NMR magnet with a field of $H_0 = 11.7294$ T, and some experiments at lower fields were conducted in a PPMS system. In this field, $T_c$ is suppressed to $\sim 5.3$ K (measured by resistivity) \cite{Vedeneev2013}. Spectra (Fig. \ref{fig:waterfall}) were collected for field aligned along the tetragonal [110], [100] and [001] directions.  The [110] ([1$\bar{1}$0]) direction corresponds to the Fe-Fe bond, and is the $\mathbf{a}$ ($\mathbf{b}$) direction in the nematic phase~\cite{Zhou2020}. The spectra were measured at several temperatures down to 2.1 K using low-power rf pulses ($\pi/2$-pulse  widths up to $80\mu$s), sweeping frequency and summing the Fourier transforms. Our results are consistent with previous reports \cite{Baek2015,Baek2016,Wiecki2017,Cao2019,Li2020}, and reveal a splitting of the single $^{77}$Se resonance below $T_{nem}  = 91$ K due to twinning.  The resonance frequencies are given by $f = \gamma H_0(1 + K)$, where $\gamma = 8.118$ MHz/T is the gyromagnetic ratio and $K$ is the Knight shift.  We fit each resonance to a Gaussian function, and Figs. \ref{fig:width}(a,b) shows the temperature dependence of $K$ and the full-width half-maxima, FWHM, for several different field directions.  Below $T_c$, the spectra exhibited a strong dependence on the pulse power, as illustrated in Fig. \ref{fig:waterfall}(c,d).  The radiofrequency pulses induce eddy currents around the sample, which can lead to Joule heating. As a result, the temperature may temporarily exceed $T_c$ immediately after the pulse.  Similar effects have been observed in other superconductors, leading to misinterpretations about the temperature dependence of the Knight shift \cite{Pustogow2019}. The shifts reported in Fig. \ref{fig:width}(b) were measured at 18.4 dBm, where there was no power-dependence to the spectra.

\begin{figure}[h]
\begin{center}
\includegraphics[width=\linewidth]{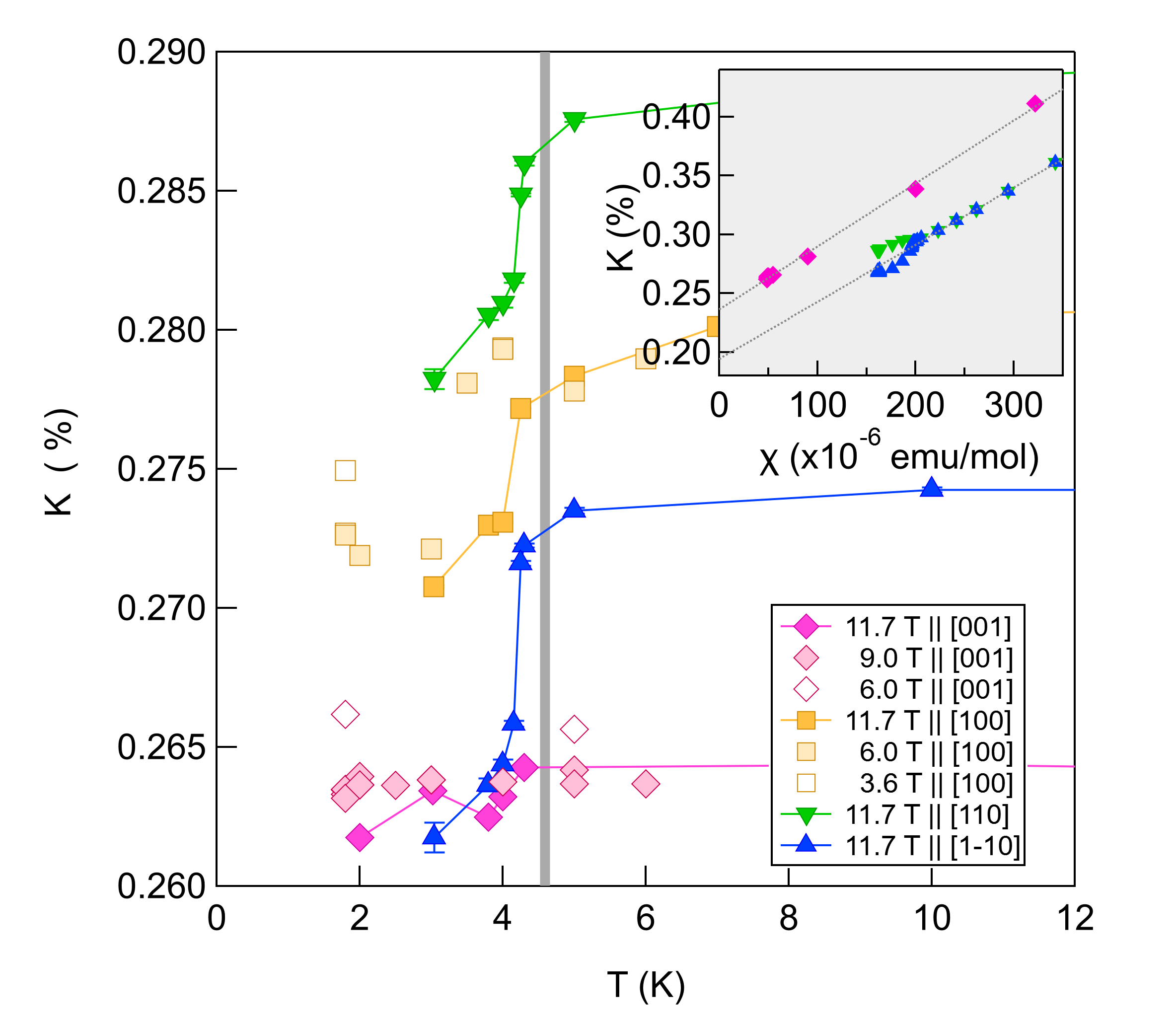}
\caption{$K_{\alpha}$ versus temperature for fields and orientations.  (INSET) $K$ versus $\chi$ for in the normal state, using susceptibility data from \cite{Li2020}.  The dotted lines indicate the best linear fits, with parameters $K_{0a} = 0.194\pm 0.002\%$, $A_{aa} =27.1\pm0.3$ kOe/$\mu_B$, $K_{0c} = 0.236\pm 0.001\%$, and $A_{cc} =29.9\pm0.5$ kOe/$\mu_B$,  as described in the text.}
\label{fig:Ksc}
\end{center}
\end{figure}

The Knight shift arises from the hyperfine interaction between the nuclear spin and the spin and orbital degrees of freedom of the electrons: $\mathcal{H}_{hf} = \mathbf{I}\cdot\mathbb{A}_S\cdot\mathbf{S} + \mathbf{I}\cdot\mathbb{A}_L\cdot\mathbf{L}$, where $\mathbb{A}_{S,L}$ are the hyperfine coupling tensors, and $\mathbf{S}$ and $\mathbf{L}$ are the spin and orbital angular momenta.  The Knight shift is given by $K_{\alpha} =  A^{L}_{\alpha\alpha}\chi_{\alpha\alpha}^{orb} +A^{S}_{\alpha\alpha}\chi_{\alpha\alpha}^{spin} + (A^{S}_{\alpha\alpha} + A^{L}_{\alpha\alpha})\chi_{\alpha\alpha}^{mixed}$, where $\chi_{\alpha\alpha}^{spin}$, $\chi_{\alpha\alpha}^{orb}$, and $\chi_{\alpha\alpha}^{mixed}$ are the static spin, orbital, and mixed susceptibilities at zero wavevector \cite{ShirerPNAS2012,Zhou2020}.   In the absence of  spin-orbit coupling, the mixed term vanishes and the Knight shift is usually decomposed as $K_{\alpha} = K_{\alpha0} + A^{S}_{\alpha\alpha}\chi_{\alpha\alpha}^{spin}$.  $K_{\alpha0}$ is often considered to be a temperature-independent shift arising from a Van-Vleck orbital susceptibility, however, this decomposition breaks down in the presence of spin-orbit coupling \cite{NissonCEFSOC2016}.  Moreover, theoretical calculations have revealed that $\chi_{\alpha\alpha}^{orb} \gg \chi_{\alpha\alpha}^{spin},\chi_{\alpha\alpha}^{mixed}$ due to the multiorbital nature of the band structure and nematic instability \cite{Zhou2020}. As a result, the relationship between $K_{\alpha}$ and the bulk susceptibility, $\chi = \chi^{spin} + \chi^{orb}+2\chi^{mixed}$, is complicated.  Nevertheless, we find that $K_{\alpha}$ varies linearly with $\chi$ above $T_{nem}$, as shown in the inset of  Fig. \ref{fig:Ksc}.   Linear fits to the data yield parameters close to previously reported values \cite{Li2020}.

For spin-singlet pairing, $\chi^{spin}$ and $\chi^{mixed}$ should vanish in the superconducting state, giving rise to a suppression of $K$ below $T_c$, as observed in Fig. \ref{fig:Ksc}.  For planar fields, $K_{a,b}$ is suppressed by about $100\pm 15$ ppm in both domains, as well as for the [100] direction oriented $45^{\circ}$ to the Fe-Fe bond direction.   This magnitude of suppression does not change significantly at lower fields.  For out of plane fields, any change in $K_c$ is within the  noise, but is less than $\sim 10$ ppm. These results are also independent of applied field, and are consistent with previous reports \cite{Kotegawa2008,Molatta2020}.  Note that $K_{\alpha} (T\rightarrow0)\neq K_{\alpha0}$, or in other words the low temperature limit of the shift does not equal the  intercepts from the $K-\chi$ plot. In fact, $\chi^{orb}$ is strongly temperature dependent, so  $K_{\alpha0}$ does not represent a temperature independent Van-Vleck term.   The low temperature shift reflects a finite $\chi^{orb}$, since the spin component vanishes for singlet pairing; however impurity states may play a role \cite{Curro2005}.   Determining how much $\chi^{spin}$, $\chi^{orb}$ and $\chi^{mixed}$ are suppressed below $T_c$ will likely require detailed theoretical calculations \cite{Zhou2020}. It is noteworthy that the difference $K_{a} - K_{b}$, shown in Fig. \ref{fig:width}(c), exhibits a subtle enhancement below $T_c$. This observation suggests that the superconductivity is slightly anisotropic in the two domains, and may reflect an anisotropy in the coherence lengths, $\xi_{a,b}$.

Below $T_c$ the spectra for both domains broaden and become asymmetric with a high frequency tail, as observed in Fig. \ref{fig:waterfall}(c,d) and \ref{fig:width}(a). At room temperature, the FWHM of the spectrum $(\sim 0.08$ kHz) is close to the second moment of the nuclear spin dipole moments of the lattice ($\sim 0.06$ kHz). The excess inhomogeneous broadening above $T_c$ may be due to either macroscopic or microscopic strain fields \cite{Tan2013}. The crystals were initially secured to the goniometer with a light coat of superglue and the coil fit loosely around the sample.  In this case, the observed linewidths were smaller (open points in Fig. \ref{fig:width}(a)), reflecting the high quality of this crystal.  In later measurements, the crystal was remounted and we observed the linewidth increase by a factor $\sim 2$ (solid points in Fig. \ref{fig:width}(a)). It is possible that remounting the crystal introduced inhomogeneous macroscopic strain fields.  Non-magnetic impurities such as Fe vacancies are known to exist in the lattice \cite{FeSeOrbitalPairing2017}, which may also be a source of  local strain and inhomogeneous broadening in the normal state.

Regardless of the linewidth in the normal state, an even larger increase of linewidth is observed below $T_c$, which is unexpected.  A vortex lattice certainly gives rise to a distribution of local magnetic fields, $B(\mathbf{r})$, and in very low fields $B\ll B_{c2}$, the second moment of the field distribution can be estimated as $\Delta B^2 \approx 0.00371\phi_0^2\lambda^{-4}$, where $\phi_0$ is the flux quantum, and the penetration depths are $\lambda_a = 446$ nm, $\lambda_c = 1320$ nm \cite{Sonier2007,AbdelHafiez2013}. There are important corrections to this expression in the higher fields where our measurements were conducted \cite{Maisuradze2009}, however after accounting for these we estimate that the normal state spectra should broaden by only $\sim 8$ Hz, three orders of magnitude smaller than the enhancement observed in Fig. \ref{fig:width}(a) \cite{supplemental}.

Since the field distribution alone is unable to capture the asymmetric broadening, we hypothesize the presence of a spatially-varying Knight shift, $K_{\alpha}(\mathbf{r})$, that is equal to the normal state value within the vortex cores and decays to $K_{\alpha}(T\rightarrow0)$ outside.  The spectrum is given by the histogram of the local resonance frequency, $f(\mathbf{r}) = \gamma B(\mathbf{r})(1 + K_{\alpha}(\mathbf{r}))$. The exact shape of the spectrum depends on microscopic details, but if the spatial variation $\delta K$ is equal to the 100 ppm suppression observed in Fig. \ref{fig:Ksc}, the spectrum will broaden by $\sim 10$ kHz, which agrees well with the excess linewidth below $T_c$ in Fig. \ref{fig:width}(a).   These results suggest that the local spin susceptibility within the vortex cores is identical to that in the normal state.

\begin{figure}[h]
\begin{center}
\includegraphics[width=\linewidth]{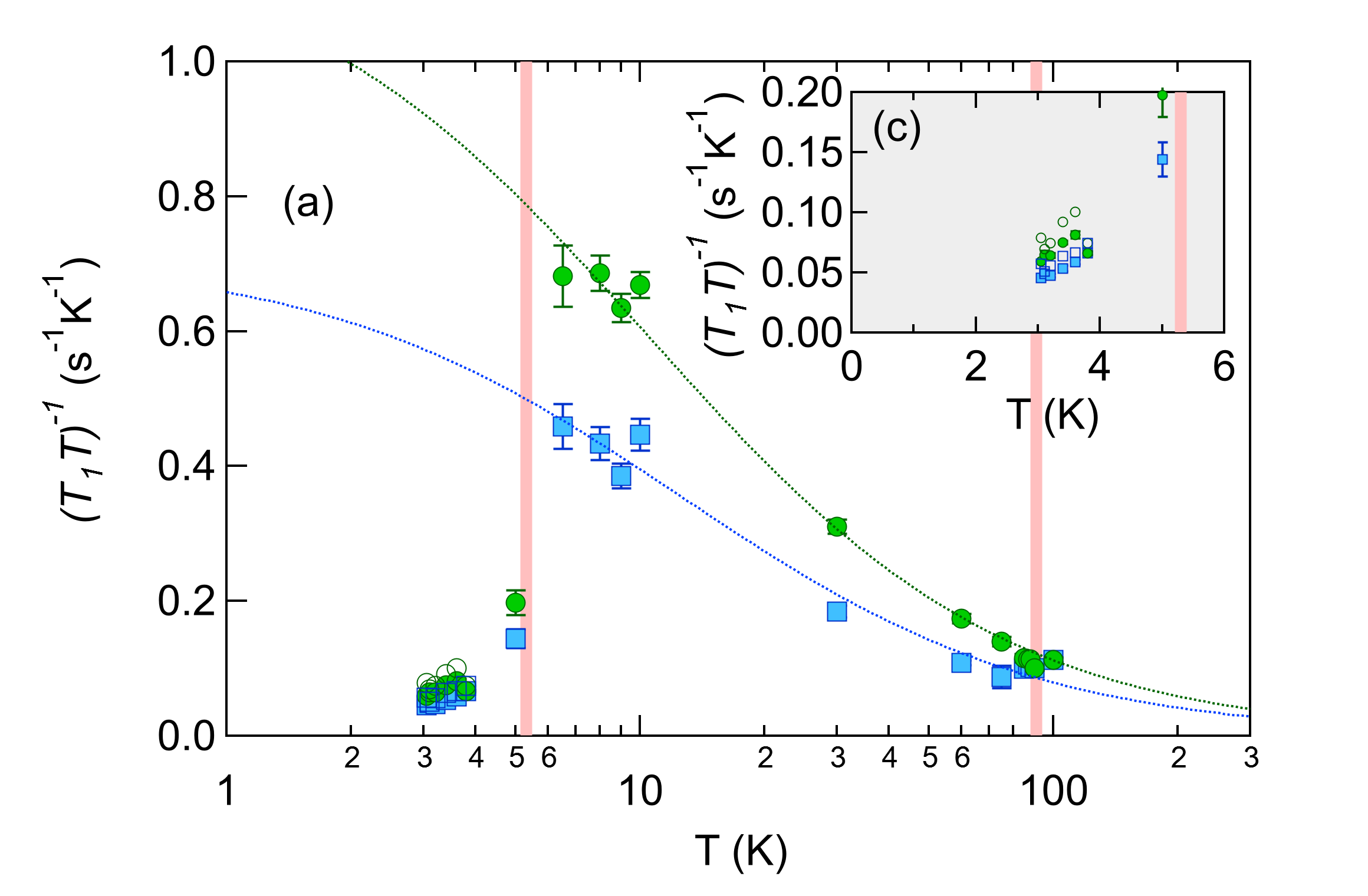}
\includegraphics[width=\linewidth]{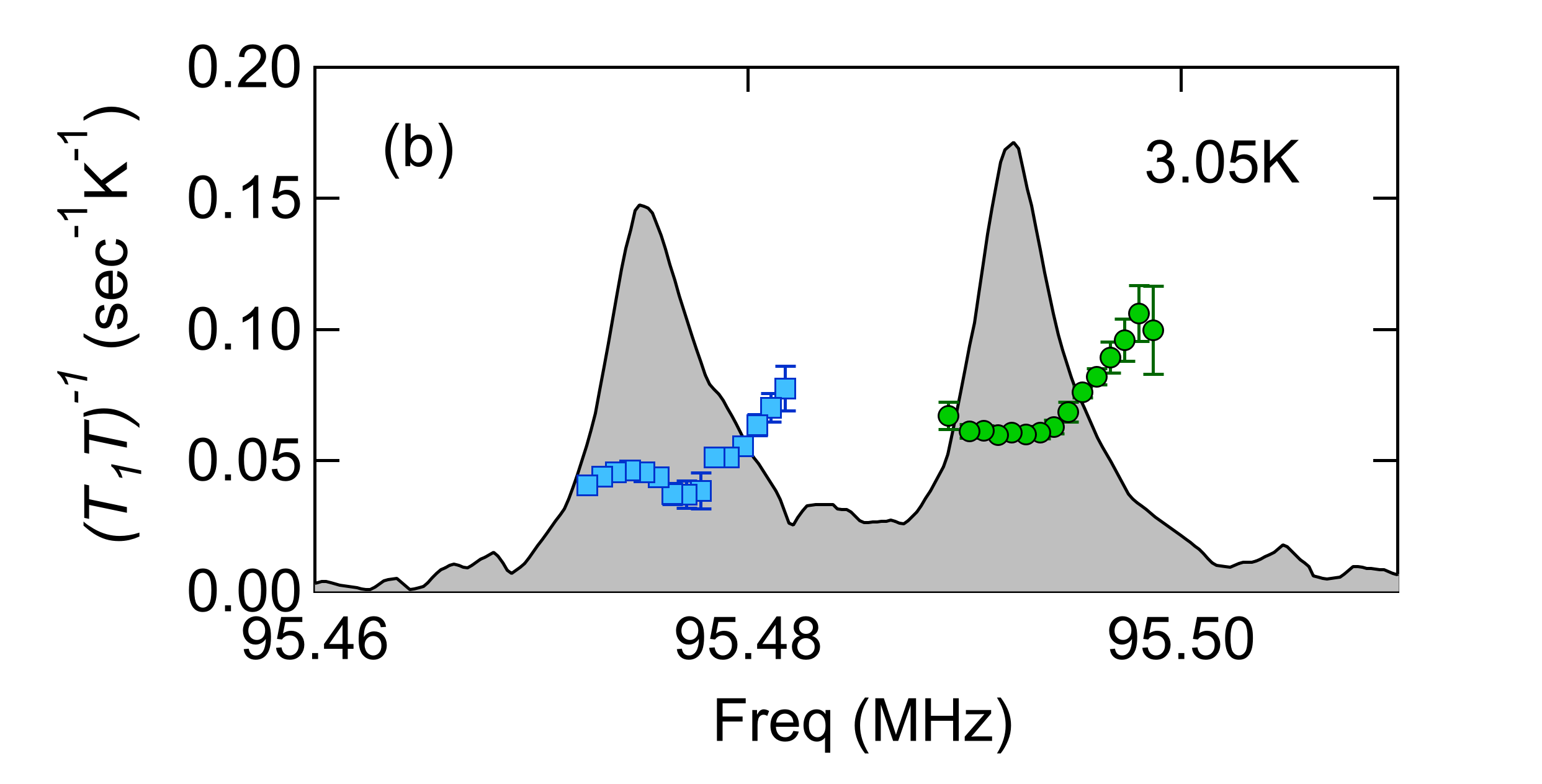}
\caption{(a) $(T_1T)^{-1}$ versus temperature (symbols defined in Fig. \ref{fig:width}).  The dotted lines are best fits to a Curie-Weiss form, as described in the text. (b) $(T_1T)^{-1}$ versus frequency at 3.05K in the mixed phase, revealing an enhanced rate within the vortex cores. The spectrum is shown in gray. \slrr\ and spectra were acquired with high power rf pulses.}
\label{fig:T1}
\end{center}
\end{figure}

This interpretation is supported by \slrr\ measurements. Fig. \ref{fig:T1}(a) shows $(T_1T)^{-1}$ versus temperature.  The data in the  normal state agree well with published results \cite{Baek2015,Shi2018,Li2020}. This quantity drops due to the superconductivity, and becomes inhomogeneous in the mixed phase. Fig. \ref{fig:T1}(b) shows that \slrr\ increases by nearly a factor of two in the high frequency tails of the spectra in the superconducting state, which correspond with the vortex cores. Localized Caroli-deGennes-Matricon (CdGM) electronic states normally exist within isolated cores \cite{Caroli1964}. At higher fields quasiparticles from different cores can propagate coherently across multiple vortices, and the energy spectrum becomes dispersive, with gapless excitations remaining within the vortex cores that give rise to a finite local density of states (LDOS) which should be manifest in any technique sensitive to low energy excitations \cite{Dukan1994,Norman1995,Ichioka1999,Zhu2016}. Indeed NMR studies have identified enhanced \slrrtext\ within the vortex cores of both conventional \cite{Nakai2008a} and unconventional superconductors  \cite{AFMinVortexMitrovicPRB2003,mitrovicYBCO}.

There are, however, important differences between FeSe and previous observations on other superconductors. In the cuprates, the excess relaxation rate has been attributed to antiferromagnetic fluctuations from a competing ground state to superconductivity \cite{AFMinVortexMitrovicPRB2003,mitrovicYBCO}, as well as from Doppler-shifted quasiparticles associated with d-wave nodes \cite{Curro2000c}.  In such cases $(T_1T)^{-1}$ exhibits a strong Curie-Weiss divergence within the cores, whereas outside the cores $(T_1T)^{-1}$ remains temperature independent. In the s-wave superconductor LaRu$_4$P$_{12}$, $(T_1T)^{-1}$ in the cores is also strongly temperature dependent, and even exceeds the value in the normal state \cite{Nakai2008a}. In the case of FeSe, $(T_1T)^{-1}$ exhibits Curie-Weiss behavior in the normal state (dotted lines in Fig. \ref{fig:T1}a), but drops below $T_c$. This behavior has been attributed to antiferromagnetic spin fluctuations that are gapped by the superconductivity \cite{Imai2009,Mukherjee2015}. The open circles (squares) in Fig. \ref{fig:T1}(a,c) show the temperature dependence at the upper end of the spectra in the superconducting state.  $(T_1T)^{-1}$ in the vortex cores changes only by a factor of two from the background rate, remaining well below the normal state value, and exhibits the same trend with temperature as the background.  These results suggest the absence of any spin fluctuations within the normal cores of FeSe. 

FeSe appears unique in that there is a $\mathbf{q}=0$ spin response in the vortex cores.  It is unclear whether this behavior could be related to either a proximity to the BCS-BEC crossover, or either a Fulde-Ferrell-Larkin-Ovchinnikov (FFLO) phase \cite{Shibauchi2020,Kasahara2020} or a field-induced spin density wave \cite{ZhouFeSeDisorder2021} for parallel fields above $H^* =24$ T. A true FFLO phase should exhibit both segmented vortex lines and normal planes where the LDOS reaches the normal state values, giving rise to inhomogeneously broadened NMR spectra. Although $H_0 \sim 0.5 H^*$ in our experiments, the inhomogeneity we observe already indicates the presence of large spin polarization in spatial regions where the superconducting order vanishes.  It is  noteworthy that the $\mathbf{q}=0$ susceptibility in FeSe is dominated by orbital contributions, whereas the finite $\mathbf{q}$ response is dominated by spin fluctuations \cite{Zhou2020}.  Condensation of singlet pairs enables us to probe the small spin response at $\mathbf{q}=0$. In Ba(Fe,Co)$_2$As$_2$, finite $\mathbf{q}$ spin fluctuations can freeze and exhibit long range antiferromagnetism in vortex cores \cite{Larsen2015}.  In FeSe we see no evidence for such behavior, which may be due to the presence of nematic order and the different contribution of orbital versus spin susceptibility. The absence of such fluctuations suggests that the high field phase is unrelated to SDW order \cite{ZhouFeSeDisorder2021,Young2007}.

In summary, we find that the Knight shift is suppressed below $T_c$ for in-plane fields, but see little to no suppression for field along the $c$-axis.  The spectra are inhomogeneously broadened below $T_c$, and \slrr\ becomes frequency-dependent.  These observations are consistent with a finite LDOS within the vortex cores. We find no evidence of competing antiferromagnetic fluctuations in the vortex cores.  Further studies at higher fields or with Te doping should shed light on the unusual nature of the  vortex states in this system.

\textit{Acknowledgment.} We acknowledge helpful discussions with B. Andersen, E. da Silva Neto, M. Walker, and R. Fernandes, and thank P. Klavins for assistance in the lab. Work at UC Davis was supported by the NSF under Grants No. DMR-1807889, and synthesis of single crystals was supported by the UC-Lab fees program.

\bibliography{FeSeNMRbibliography}

\begin{thebibliography}{51}%
\makeatletter
\providecommand \@ifxundefined [1]{%
 \@ifx{#1\undefined}
}%
\providecommand \@ifnum [1]{%
 \ifnum #1\expandafter \@firstoftwo
 \else \expandafter \@secondoftwo
 \fi
}%
\providecommand \@ifx [1]{%
 \ifx #1\expandafter \@firstoftwo
 \else \expandafter \@secondoftwo
 \fi
}%
\providecommand \natexlab [1]{#1}%
\providecommand \enquote  [1]{``#1''}%
\providecommand \bibnamefont  [1]{#1}%
\providecommand \bibfnamefont [1]{#1}%
\providecommand \citenamefont [1]{#1}%
\providecommand \href@noop [0]{\@secondoftwo}%
\providecommand \href [0]{\begingroup \@sanitize@url \@href}%
\providecommand \@href[1]{\@@startlink{#1}\@@href}%
\providecommand \@@href[1]{\endgroup#1\@@endlink}%
\providecommand \@sanitize@url [0]{\catcode `\\12\catcode `\$12\catcode
  `\&12\catcode `\#12\catcode `\^12\catcode `\_12\catcode `\%12\relax}%
\providecommand \@@startlink[1]{}%
\providecommand \@@endlink[0]{}%
\providecommand \url  [0]{\begingroup\@sanitize@url \@url }%
\providecommand \@url [1]{\endgroup\@href {#1}{\urlprefix }}%
\providecommand \urlprefix  [0]{URL }%
\providecommand \Eprint [0]{\href }%
\providecommand \doibase [0]{https://doi.org/}%
\providecommand \selectlanguage [0]{\@gobble}%
\providecommand \bibinfo  [0]{\@secondoftwo}%
\providecommand \bibfield  [0]{\@secondoftwo}%
\providecommand \translation [1]{[#1]}%
\providecommand \BibitemOpen [0]{}%
\providecommand \bibitemStop [0]{}%
\providecommand \bibitemNoStop [0]{.\EOS\space}%
\providecommand \EOS [0]{\spacefactor3000\relax}%
\providecommand \BibitemShut  [1]{\csname bibitem#1\endcsname}%
\let\auto@bib@innerbib\@empty
\bibitem [{\citenamefont {K\"{o}nig}\ and\ \citenamefont
  {Coleman}(2019)}]{Koenig2019}%
  \BibitemOpen
  \bibfield  {author} {\bibinfo {author} {\bibfnamefont {E.~J.}\ \bibnamefont
  {K\"{o}nig}}\ and\ \bibinfo {author} {\bibfnamefont {P.}~\bibnamefont
  {Coleman}},\ }\bibfield  {title} {\bibinfo {title}
  {Crystalline-symmetry-protected helical {M}ajorana modes in the iron
  pnictides},\ }\href {https://doi.org/10.1103/physrevlett.122.207001}
  {\bibfield  {journal} {\bibinfo  {journal} {Phys. Rev. Lett.}\ }\textbf
  {\bibinfo {volume} {122}},\ \bibinfo {pages} {207001} (\bibinfo {year}
  {2019})}\BibitemShut {NoStop}%
\bibitem [{\citenamefont {Zhang}\ \emph {et~al.}(2018)\citenamefont {Zhang},
  \citenamefont {Wang}, \citenamefont {Wu}, \citenamefont {Yaji}, \citenamefont
  {Ishida}, \citenamefont {Kohama}, \citenamefont {Dai}, \citenamefont {Sun},
  \citenamefont {Bareille}, \citenamefont {Kuroda}, \citenamefont {Kondo},
  \citenamefont {Okazaki}, \citenamefont {Kindo}, \citenamefont {Wang},
  \citenamefont {Jin}, \citenamefont {Hu}, \citenamefont {Thomale},
  \citenamefont {Sumida}, \citenamefont {Wu}, \citenamefont {Miyamoto},
  \citenamefont {Okuda}, \citenamefont {Ding}, \citenamefont {Gu},
  \citenamefont {Tamegai}, \citenamefont {Kawakami}, \citenamefont {Sato},\
  and\ \citenamefont {Shin}}]{Zhang2018}%
  \BibitemOpen
  \bibfield  {author} {\bibinfo {author} {\bibfnamefont {P.}~\bibnamefont
  {Zhang}}, \bibinfo {author} {\bibfnamefont {Z.}~\bibnamefont {Wang}},
  \bibinfo {author} {\bibfnamefont {X.}~\bibnamefont {Wu}}, \bibinfo {author}
  {\bibfnamefont {K.}~\bibnamefont {Yaji}}, \bibinfo {author} {\bibfnamefont
  {Y.}~\bibnamefont {Ishida}}, \bibinfo {author} {\bibfnamefont
  {Y.}~\bibnamefont {Kohama}}, \bibinfo {author} {\bibfnamefont
  {G.}~\bibnamefont {Dai}}, \bibinfo {author} {\bibfnamefont {Y.}~\bibnamefont
  {Sun}}, \bibinfo {author} {\bibfnamefont {C.}~\bibnamefont {Bareille}},
  \bibinfo {author} {\bibfnamefont {K.}~\bibnamefont {Kuroda}}, \bibinfo
  {author} {\bibfnamefont {T.}~\bibnamefont {Kondo}}, \bibinfo {author}
  {\bibfnamefont {K.}~\bibnamefont {Okazaki}}, \bibinfo {author} {\bibfnamefont
  {K.}~\bibnamefont {Kindo}}, \bibinfo {author} {\bibfnamefont
  {X.}~\bibnamefont {Wang}}, \bibinfo {author} {\bibfnamefont {C.}~\bibnamefont
  {Jin}}, \bibinfo {author} {\bibfnamefont {J.}~\bibnamefont {Hu}}, \bibinfo
  {author} {\bibfnamefont {R.}~\bibnamefont {Thomale}}, \bibinfo {author}
  {\bibfnamefont {K.}~\bibnamefont {Sumida}}, \bibinfo {author} {\bibfnamefont
  {S.}~\bibnamefont {Wu}}, \bibinfo {author} {\bibfnamefont {K.}~\bibnamefont
  {Miyamoto}}, \bibinfo {author} {\bibfnamefont {T.}~\bibnamefont {Okuda}},
  \bibinfo {author} {\bibfnamefont {H.}~\bibnamefont {Ding}}, \bibinfo {author}
  {\bibfnamefont {G.~D.}\ \bibnamefont {Gu}}, \bibinfo {author} {\bibfnamefont
  {T.}~\bibnamefont {Tamegai}}, \bibinfo {author} {\bibfnamefont
  {T.}~\bibnamefont {Kawakami}}, \bibinfo {author} {\bibfnamefont
  {M.}~\bibnamefont {Sato}},\ and\ \bibinfo {author} {\bibfnamefont
  {S.}~\bibnamefont {Shin}},\ }\bibfield  {title} {\bibinfo {title} {Multiple
  topological states in iron-based superconductors},\ }\href
  {https://doi.org/10.1038/s41567-018-0280-z} {\bibfield  {journal} {\bibinfo
  {journal} {Nat. Phys.}\ }\textbf {\bibinfo {volume} {15}},\ \bibinfo {pages}
  {41} (\bibinfo {year} {2018})}\BibitemShut {NoStop}%
\bibitem [{\citenamefont {{Kong}}\ \emph {et~al.}(2020)\citenamefont {{Kong}},
  \citenamefont {{Cao}}, \citenamefont {{Zhu}}, \citenamefont {{Papaj}},
  \citenamefont {{Dai}}, \citenamefont {{Li}}, \citenamefont {{Fan}},
  \citenamefont {{Liu}}, \citenamefont {{Yang}}, \citenamefont {{Wang}},
  \citenamefont {{Du}}, \citenamefont {{Jin}}, \citenamefont {{Fu}},
  \citenamefont {{Gao}},\ and\ \citenamefont {{Ding}}}]{Kong2020}%
  \BibitemOpen
  \bibfield  {author} {\bibinfo {author} {\bibfnamefont {L.}~\bibnamefont
  {{Kong}}}, \bibinfo {author} {\bibfnamefont {L.}~\bibnamefont {{Cao}}},
  \bibinfo {author} {\bibfnamefont {S.}~\bibnamefont {{Zhu}}}, \bibinfo
  {author} {\bibfnamefont {M.}~\bibnamefont {{Papaj}}}, \bibinfo {author}
  {\bibfnamefont {G.}~\bibnamefont {{Dai}}}, \bibinfo {author} {\bibfnamefont
  {G.}~\bibnamefont {{Li}}}, \bibinfo {author} {\bibfnamefont {P.}~\bibnamefont
  {{Fan}}}, \bibinfo {author} {\bibfnamefont {W.}~\bibnamefont {{Liu}}},
  \bibinfo {author} {\bibfnamefont {F.}~\bibnamefont {{Yang}}}, \bibinfo
  {author} {\bibfnamefont {X.}~\bibnamefont {{Wang}}}, \bibinfo {author}
  {\bibfnamefont {S.}~\bibnamefont {{Du}}}, \bibinfo {author} {\bibfnamefont
  {C.}~\bibnamefont {{Jin}}}, \bibinfo {author} {\bibfnamefont
  {L.}~\bibnamefont {{Fu}}}, \bibinfo {author} {\bibfnamefont {H.-J.}\
  \bibnamefont {{Gao}}},\ and\ \bibinfo {author} {\bibfnamefont
  {H.}~\bibnamefont {{Ding}}},\ }\bibfield  {title} {\bibinfo {title} {{Tunable
  vortex Majorana zero modes in LiFeAs superconductor}},\ }\href@noop {}
  {\bibfield  {journal} {\bibinfo  {journal} {arXiv e-prints}\ ,\ \bibinfo
  {eid} {arXiv:2010.04735}} (\bibinfo {year} {2020})},\ \Eprint
  {https://arxiv.org/abs/2010.04735} {arXiv:2010.04735 [cond-mat.supr-con]}
  \BibitemShut {NoStop}%
\bibitem [{\citenamefont {Wang}\ \emph {et~al.}(2020)\citenamefont {Wang},
  \citenamefont {Rodriguez}, \citenamefont {Jiao}, \citenamefont {Howard},
  \citenamefont {Graham}, \citenamefont {Gu}, \citenamefont {Hughes},
  \citenamefont {Morr},\ and\ \citenamefont {Madhavan}}]{Wang2020}%
  \BibitemOpen
  \bibfield  {author} {\bibinfo {author} {\bibfnamefont {Z.}~\bibnamefont
  {Wang}}, \bibinfo {author} {\bibfnamefont {J.~O.}\ \bibnamefont {Rodriguez}},
  \bibinfo {author} {\bibfnamefont {L.}~\bibnamefont {Jiao}}, \bibinfo {author}
  {\bibfnamefont {S.}~\bibnamefont {Howard}}, \bibinfo {author} {\bibfnamefont
  {M.}~\bibnamefont {Graham}}, \bibinfo {author} {\bibfnamefont {G.~D.}\
  \bibnamefont {Gu}}, \bibinfo {author} {\bibfnamefont {T.~L.}\ \bibnamefont
  {Hughes}}, \bibinfo {author} {\bibfnamefont {D.~K.}\ \bibnamefont {Morr}},\
  and\ \bibinfo {author} {\bibfnamefont {V.}~\bibnamefont {Madhavan}},\
  }\bibfield  {title} {\bibinfo {title} {Evidence for dispersing 1d majorana
  channels in an iron-based superconductor},\ }\href
  {https://doi.org/10.1126/science.aaw8419} {\bibfield  {journal} {\bibinfo
  {journal} {Science}\ }\textbf {\bibinfo {volume} {367}},\ \bibinfo {pages}
  {104} (\bibinfo {year} {2020})}\BibitemShut {NoStop}%
\bibitem [{\citenamefont {Wang}\ \emph {et~al.}(2015)\citenamefont {Wang},
  \citenamefont {Zhang}, \citenamefont {Xu}, \citenamefont {Zeng},
  \citenamefont {Miao}, \citenamefont {Xu}, \citenamefont {Qian}, \citenamefont
  {Weng}, \citenamefont {Richard}, \citenamefont {Fedorov}, \citenamefont
  {Ding}, \citenamefont {Dai},\ and\ \citenamefont {Fang}}]{Wang2015}%
  \BibitemOpen
  \bibfield  {author} {\bibinfo {author} {\bibfnamefont {Z.}~\bibnamefont
  {Wang}}, \bibinfo {author} {\bibfnamefont {P.}~\bibnamefont {Zhang}},
  \bibinfo {author} {\bibfnamefont {G.}~\bibnamefont {Xu}}, \bibinfo {author}
  {\bibfnamefont {L.~K.}\ \bibnamefont {Zeng}}, \bibinfo {author}
  {\bibfnamefont {H.}~\bibnamefont {Miao}}, \bibinfo {author} {\bibfnamefont
  {X.}~\bibnamefont {Xu}}, \bibinfo {author} {\bibfnamefont {T.}~\bibnamefont
  {Qian}}, \bibinfo {author} {\bibfnamefont {H.}~\bibnamefont {Weng}}, \bibinfo
  {author} {\bibfnamefont {P.}~\bibnamefont {Richard}}, \bibinfo {author}
  {\bibfnamefont {A.~V.}\ \bibnamefont {Fedorov}}, \bibinfo {author}
  {\bibfnamefont {H.}~\bibnamefont {Ding}}, \bibinfo {author} {\bibfnamefont
  {X.}~\bibnamefont {Dai}},\ and\ \bibinfo {author} {\bibfnamefont
  {Z.}~\bibnamefont {Fang}},\ }\bibfield  {title} {\bibinfo {title}
  {Topological nature of the {FeSe$_{0.5}$Te$_{0.5}$} superconductor},\ }\href
  {https://doi.org/10.1103/physrevb.92.115119} {\bibfield  {journal} {\bibinfo
  {journal} {Phys. Rev. B}\ }\textbf {\bibinfo {volume} {92}},\ \bibinfo
  {pages} {115119} (\bibinfo {year} {2015})}\BibitemShut {NoStop}%
\bibitem [{\citenamefont {Kreisel}\ \emph {et~al.}(2020)\citenamefont
  {Kreisel}, \citenamefont {Hirschfeld},\ and\ \citenamefont
  {Andersen}}]{Kreisel2020}%
  \BibitemOpen
  \bibfield  {author} {\bibinfo {author} {\bibfnamefont {A.}~\bibnamefont
  {Kreisel}}, \bibinfo {author} {\bibfnamefont {P.~J.}\ \bibnamefont
  {Hirschfeld}},\ and\ \bibinfo {author} {\bibfnamefont {B.~M.}\ \bibnamefont
  {Andersen}},\ }\bibfield  {title} {\bibinfo {title} {On the remarkable
  superconductivity of {FeSe} and its close cousins},\ }\href
  {https://doi.org/10.3390/sym12091402} {\bibfield  {journal} {\bibinfo
  {journal} {Symmetry}\ }\textbf {\bibinfo {volume} {12}},\ \bibinfo {pages}
  {1402} (\bibinfo {year} {2020})}\BibitemShut {NoStop}%
\bibitem [{\citenamefont {Hanaguri}\ \emph {et~al.}(2019)\citenamefont
  {Hanaguri}, \citenamefont {Kasahara}, \citenamefont {Böker}, \citenamefont
  {Eremin}, \citenamefont {Shibauchi},\ and\ \citenamefont
  {Matsuda}}]{Hanaguri2019}%
  \BibitemOpen
  \bibfield  {author} {\bibinfo {author} {\bibfnamefont {T.}~\bibnamefont
  {Hanaguri}}, \bibinfo {author} {\bibfnamefont {S.}~\bibnamefont {Kasahara}},
  \bibinfo {author} {\bibfnamefont {J.}~\bibnamefont {Böker}}, \bibinfo
  {author} {\bibfnamefont {I.}~\bibnamefont {Eremin}}, \bibinfo {author}
  {\bibfnamefont {T.}~\bibnamefont {Shibauchi}},\ and\ \bibinfo {author}
  {\bibfnamefont {Y.}~\bibnamefont {Matsuda}},\ }\bibfield  {title} {\bibinfo
  {title} {Quantum vortex core and missing pseudogap in the multiband
  {BCS}-{BEC} crossover superconductor {FeSe}},\ }\href
  {https://doi.org/10.1103/physrevlett.122.077001} {\bibfield  {journal}
  {\bibinfo  {journal} {Phys. Rev. Lett.}\ }\textbf {\bibinfo {volume} {122}},\
  \bibinfo {pages} {077001} (\bibinfo {year} {2019})}\BibitemShut {NoStop}%
\bibitem [{\citenamefont {Shibauchi}\ \emph {et~al.}(2020)\citenamefont
  {Shibauchi}, \citenamefont {Hanaguri},\ and\ \citenamefont
  {Matsuda}}]{Shibauchi2020}%
  \BibitemOpen
  \bibfield  {author} {\bibinfo {author} {\bibfnamefont {T.}~\bibnamefont
  {Shibauchi}}, \bibinfo {author} {\bibfnamefont {T.}~\bibnamefont
  {Hanaguri}},\ and\ \bibinfo {author} {\bibfnamefont {Y.}~\bibnamefont
  {Matsuda}},\ }\bibfield  {title} {\bibinfo {title} {Exotic superconducting
  states in {FeSe}-based materials},\ }\href
  {https://doi.org/10.7566/jpsj.89.102002} {\bibfield  {journal} {\bibinfo
  {journal} {J. Phys. Soc. Japan}\ }\textbf {\bibinfo {volume} {89}},\ \bibinfo
  {pages} {102002} (\bibinfo {year} {2020})}\BibitemShut {NoStop}%
\bibitem [{\citenamefont {Kasahara}\ \emph {et~al.}(2020)\citenamefont
  {Kasahara}, \citenamefont {Sato}, \citenamefont {Licciardello}, \citenamefont
  {{\v{C}}ulo}, \citenamefont {Arsenijevi{\'{c}}}, \citenamefont {Ottenbros},
  \citenamefont {Tominaga}, \citenamefont {Böker}, \citenamefont {Eremin},
  \citenamefont {Shibauchi}, \citenamefont {Wosnitza}, \citenamefont {Hussey},\
  and\ \citenamefont {Matsuda}}]{Kasahara2020}%
  \BibitemOpen
  \bibfield  {author} {\bibinfo {author} {\bibfnamefont {S.}~\bibnamefont
  {Kasahara}}, \bibinfo {author} {\bibfnamefont {Y.}~\bibnamefont {Sato}},
  \bibinfo {author} {\bibfnamefont {S.}~\bibnamefont {Licciardello}}, \bibinfo
  {author} {\bibfnamefont {M.}~\bibnamefont {{\v{C}}ulo}}, \bibinfo {author}
  {\bibfnamefont {S.}~\bibnamefont {Arsenijevi{\'{c}}}}, \bibinfo {author}
  {\bibfnamefont {T.}~\bibnamefont {Ottenbros}}, \bibinfo {author}
  {\bibfnamefont {T.}~\bibnamefont {Tominaga}}, \bibinfo {author}
  {\bibfnamefont {J.}~\bibnamefont {Böker}}, \bibinfo {author} {\bibfnamefont
  {I.}~\bibnamefont {Eremin}}, \bibinfo {author} {\bibfnamefont
  {T.}~\bibnamefont {Shibauchi}}, \bibinfo {author} {\bibfnamefont
  {J.}~\bibnamefont {Wosnitza}}, \bibinfo {author} {\bibfnamefont
  {N.}~\bibnamefont {Hussey}},\ and\ \bibinfo {author} {\bibfnamefont
  {Y.}~\bibnamefont {Matsuda}},\ }\bibfield  {title} {\bibinfo {title}
  {Evidence for an fulde-ferrell-larkin-ovchinnikov state with segmented
  vortices in the {BCS}-{BEC}-crossover superconductor {FeSe}},\ }\href
  {https://doi.org/10.1103/physrevlett.124.107001} {\bibfield  {journal}
  {\bibinfo  {journal} {Phys. Rev. Lett.}\ }\textbf {\bibinfo {volume} {124}},\
  \bibinfo {pages} {107001} (\bibinfo {year} {2020})}\BibitemShut {NoStop}%
\bibitem [{\citenamefont {Sprau}\ \emph {et~al.}(2017)\citenamefont {Sprau},
  \citenamefont {Kostin}, \citenamefont {Kreisel}, \citenamefont {B{\"o}hmer},
  \citenamefont {Taufour}, \citenamefont {Canfield}, \citenamefont {Mukherjee},
  \citenamefont {Hirschfeld}, \citenamefont {Andersen},\ and\ \citenamefont
  {Davis}}]{FeSeOrbitalPairing2017}%
  \BibitemOpen
  \bibfield  {author} {\bibinfo {author} {\bibfnamefont {P.~O.}\ \bibnamefont
  {Sprau}}, \bibinfo {author} {\bibfnamefont {A.}~\bibnamefont {Kostin}},
  \bibinfo {author} {\bibfnamefont {A.}~\bibnamefont {Kreisel}}, \bibinfo
  {author} {\bibfnamefont {A.~E.}\ \bibnamefont {B{\"o}hmer}}, \bibinfo
  {author} {\bibfnamefont {V.}~\bibnamefont {Taufour}}, \bibinfo {author}
  {\bibfnamefont {P.~C.}\ \bibnamefont {Canfield}}, \bibinfo {author}
  {\bibfnamefont {S.}~\bibnamefont {Mukherjee}}, \bibinfo {author}
  {\bibfnamefont {P.~J.}\ \bibnamefont {Hirschfeld}}, \bibinfo {author}
  {\bibfnamefont {B.~M.}\ \bibnamefont {Andersen}},\ and\ \bibinfo {author}
  {\bibfnamefont {J.~C.~S.}\ \bibnamefont {Davis}},\ }\bibfield  {title}
  {\bibinfo {title} {Discovery of orbital-selective cooper pairing in {FeSe}},\
  }\href {https://doi.org/10.1126/science.aal1575} {\bibfield  {journal}
  {\bibinfo  {journal} {Science}\ }\textbf {\bibinfo {volume} {357}},\ \bibinfo
  {pages} {75} (\bibinfo {year} {2017})}\BibitemShut {NoStop}%
\bibitem [{\citenamefont {Bourgeois-Hope}\ \emph {et~al.}(2016)\citenamefont
  {Bourgeois-Hope}, \citenamefont {Chi}, \citenamefont {Bonn}, \citenamefont
  {Liang}, \citenamefont {Hardy}, \citenamefont {Wolf}, \citenamefont
  {Meingast}, \citenamefont {Doiron-Leyraud},\ and\ \citenamefont
  {Taillefer}}]{BourgeoisHope2016}%
  \BibitemOpen
  \bibfield  {author} {\bibinfo {author} {\bibfnamefont {P.}~\bibnamefont
  {Bourgeois-Hope}}, \bibinfo {author} {\bibfnamefont {S.}~\bibnamefont {Chi}},
  \bibinfo {author} {\bibfnamefont {D.}~\bibnamefont {Bonn}}, \bibinfo {author}
  {\bibfnamefont {R.}~\bibnamefont {Liang}}, \bibinfo {author} {\bibfnamefont
  {W.}~\bibnamefont {Hardy}}, \bibinfo {author} {\bibfnamefont
  {T.}~\bibnamefont {Wolf}}, \bibinfo {author} {\bibfnamefont {C.}~\bibnamefont
  {Meingast}}, \bibinfo {author} {\bibfnamefont {N.}~\bibnamefont
  {Doiron-Leyraud}},\ and\ \bibinfo {author} {\bibfnamefont {L.}~\bibnamefont
  {Taillefer}},\ }\bibfield  {title} {\bibinfo {title} {Thermal conductivity of
  the iron-based superconductor {FeSe}: Nodeless gap with a strong two-band
  character},\ }\href {https://doi.org/10.1103/physrevlett.117.097003}
  {\bibfield  {journal} {\bibinfo  {journal} {Phys. Rev. Lett.}\ }\textbf
  {\bibinfo {volume} {117}},\ \bibinfo {pages} {097003} (\bibinfo {year}
  {2016})}\BibitemShut {NoStop}%
\bibitem [{\citenamefont {Jiao}\ \emph {et~al.}(2017)\citenamefont {Jiao},
  \citenamefont {Huang}, \citenamefont {Rö{\ss}ler}, \citenamefont {Koz},
  \citenamefont {Rö{\ss}ler}, \citenamefont {Schwarz},\ and\ \citenamefont
  {Wirth}}]{Jiao2017}%
  \BibitemOpen
  \bibfield  {author} {\bibinfo {author} {\bibfnamefont {L.}~\bibnamefont
  {Jiao}}, \bibinfo {author} {\bibfnamefont {C.-L.}\ \bibnamefont {Huang}},
  \bibinfo {author} {\bibfnamefont {S.}~\bibnamefont {Rö{\ss}ler}}, \bibinfo
  {author} {\bibfnamefont {C.}~\bibnamefont {Koz}}, \bibinfo {author}
  {\bibfnamefont {U.~K.}\ \bibnamefont {Rö{\ss}ler}}, \bibinfo {author}
  {\bibfnamefont {U.}~\bibnamefont {Schwarz}},\ and\ \bibinfo {author}
  {\bibfnamefont {S.}~\bibnamefont {Wirth}},\ }\bibfield  {title} {\bibinfo
  {title} {Superconducting gap structure of {FeSe}},\ }\href
  {https://doi.org/10.1038/srep44024} {\bibfield  {journal} {\bibinfo
  {journal} {Sci. Rep.}\ }\textbf {\bibinfo {volume} {7}},\ \bibinfo {pages}
  {44024} (\bibinfo {year} {2017})}\BibitemShut {NoStop}%
\bibitem [{\citenamefont {Biswas}\ \emph {et~al.}(2018)\citenamefont {Biswas},
  \citenamefont {Kreisel}, \citenamefont {Wang}, \citenamefont {Adroja},
  \citenamefont {Hillier}, \citenamefont {Zhao}, \citenamefont {Khasanov},
  \citenamefont {Orain}, \citenamefont {Amato},\ and\ \citenamefont
  {Morenzoni}}]{Biswas2018}%
  \BibitemOpen
  \bibfield  {author} {\bibinfo {author} {\bibfnamefont {P.~K.}\ \bibnamefont
  {Biswas}}, \bibinfo {author} {\bibfnamefont {A.}~\bibnamefont {Kreisel}},
  \bibinfo {author} {\bibfnamefont {Q.}~\bibnamefont {Wang}}, \bibinfo {author}
  {\bibfnamefont {D.~T.}\ \bibnamefont {Adroja}}, \bibinfo {author}
  {\bibfnamefont {A.~D.}\ \bibnamefont {Hillier}}, \bibinfo {author}
  {\bibfnamefont {J.}~\bibnamefont {Zhao}}, \bibinfo {author} {\bibfnamefont
  {R.}~\bibnamefont {Khasanov}}, \bibinfo {author} {\bibfnamefont {J.-C.}\
  \bibnamefont {Orain}}, \bibinfo {author} {\bibfnamefont {A.}~\bibnamefont
  {Amato}},\ and\ \bibinfo {author} {\bibfnamefont {E.}~\bibnamefont
  {Morenzoni}},\ }\bibfield  {title} {\bibinfo {title} {Evidence of nodal gap
  structure in the basal plane of the {FeSe} superconductor},\ }\href
  {https://doi.org/10.1103/PhysRevB.98.180501} {\bibfield  {journal} {\bibinfo
  {journal} {Phys. Rev. B}\ }\textbf {\bibinfo {volume} {98}},\ \bibinfo
  {pages} {180501} (\bibinfo {year} {2018})}\BibitemShut {NoStop}%
\bibitem [{\citenamefont {Hardy}\ \emph {et~al.}(2019)\citenamefont {Hardy},
  \citenamefont {He}, \citenamefont {Wang}, \citenamefont {Wolf}, \citenamefont
  {Schweiss}, \citenamefont {Merz}, \citenamefont {Barth}, \citenamefont
  {Adelmann}, \citenamefont {Eder}, \citenamefont {Haghighirad},\ and\
  \citenamefont {Meingast}}]{Hardy2019}%
  \BibitemOpen
  \bibfield  {author} {\bibinfo {author} {\bibfnamefont {F.}~\bibnamefont
  {Hardy}}, \bibinfo {author} {\bibfnamefont {M.}~\bibnamefont {He}}, \bibinfo
  {author} {\bibfnamefont {L.}~\bibnamefont {Wang}}, \bibinfo {author}
  {\bibfnamefont {T.}~\bibnamefont {Wolf}}, \bibinfo {author} {\bibfnamefont
  {P.}~\bibnamefont {Schweiss}}, \bibinfo {author} {\bibfnamefont
  {M.}~\bibnamefont {Merz}}, \bibinfo {author} {\bibfnamefont {M.}~\bibnamefont
  {Barth}}, \bibinfo {author} {\bibfnamefont {P.}~\bibnamefont {Adelmann}},
  \bibinfo {author} {\bibfnamefont {R.}~\bibnamefont {Eder}}, \bibinfo {author}
  {\bibfnamefont {A.-A.}\ \bibnamefont {Haghighirad}},\ and\ \bibinfo {author}
  {\bibfnamefont {C.}~\bibnamefont {Meingast}},\ }\bibfield  {title} {\bibinfo
  {title} {Calorimetric evidence of nodal gaps in the nematic superconductor
  {FeSe}},\ }\href {https://doi.org/10.1103/physrevb.99.035157} {\bibfield
  {journal} {\bibinfo  {journal} {Phys. Rev. B}\ }\textbf {\bibinfo {volume}
  {99}},\ \bibinfo {pages} {035157} (\bibinfo {year} {2019})}\BibitemShut
  {NoStop}%
\bibitem [{\citenamefont {Baek}\ \emph {et~al.}(2015)\citenamefont {Baek},
  \citenamefont {Efremov}, \citenamefont {Ok}, \citenamefont {Kim},
  \citenamefont {van~den Brink},\ and\ \citenamefont {B\"{u}chner}}]{Baek2015}%
  \BibitemOpen
  \bibfield  {author} {\bibinfo {author} {\bibfnamefont {S.-H.}\ \bibnamefont
  {Baek}}, \bibinfo {author} {\bibfnamefont {D.~V.}\ \bibnamefont {Efremov}},
  \bibinfo {author} {\bibfnamefont {J.~M.}\ \bibnamefont {Ok}}, \bibinfo
  {author} {\bibfnamefont {J.~S.}\ \bibnamefont {Kim}}, \bibinfo {author}
  {\bibfnamefont {J.}~\bibnamefont {van~den Brink}},\ and\ \bibinfo {author}
  {\bibfnamefont {B.}~\bibnamefont {B\"{u}chner}},\ }\bibfield  {title}
  {\bibinfo {title} {Orbital-driven nematicity in {FeSe}},\ }\href
  {http://dx.doi.org/10.1038/nmat4138} {\bibfield  {journal} {\bibinfo
  {journal} {Nat. Mater.}\ }\textbf {\bibinfo {volume} {14}},\ \bibinfo {pages}
  {210} (\bibinfo {year} {2015})}\BibitemShut {NoStop}%
\bibitem [{\citenamefont {Wiecki}\ \emph {et~al.}(2017)\citenamefont {Wiecki},
  \citenamefont {Nandi}, \citenamefont {Böhmer}, \citenamefont {Bud'ko},
  \citenamefont {Canfield},\ and\ \citenamefont {Furukawa}}]{Wiecki2017}%
  \BibitemOpen
  \bibfield  {author} {\bibinfo {author} {\bibfnamefont {P.}~\bibnamefont
  {Wiecki}}, \bibinfo {author} {\bibfnamefont {M.}~\bibnamefont {Nandi}},
  \bibinfo {author} {\bibfnamefont {A.~E.}\ \bibnamefont {Böhmer}}, \bibinfo
  {author} {\bibfnamefont {S.~L.}\ \bibnamefont {Bud'ko}}, \bibinfo {author}
  {\bibfnamefont {P.~C.}\ \bibnamefont {Canfield}},\ and\ \bibinfo {author}
  {\bibfnamefont {Y.}~\bibnamefont {Furukawa}},\ }\bibfield  {title} {\bibinfo
  {title} {{NMR} evidence for static local nematicity and its cooperative
  interplay with low-energy magnetic fluctuations in {FeSe} under pressure},\
  }\href {https://doi.org/10.1103/physrevb.96.180502} {\bibfield  {journal}
  {\bibinfo  {journal} {Phys. Rev. B}\ }\textbf {\bibinfo {volume} {96}},\
  \bibinfo {pages} {180502} (\bibinfo {year} {2017})}\BibitemShut {NoStop}%
\bibitem [{\citenamefont {Baek}\ \emph {et~al.}(2016)\citenamefont {Baek},
  \citenamefont {Efremov}, \citenamefont {Ok}, \citenamefont {Kim},
  \citenamefont {van~den Brink},\ and\ \citenamefont {Büchner}}]{Baek2016}%
  \BibitemOpen
  \bibfield  {author} {\bibinfo {author} {\bibfnamefont {S.-H.}\ \bibnamefont
  {Baek}}, \bibinfo {author} {\bibfnamefont {D.~V.}\ \bibnamefont {Efremov}},
  \bibinfo {author} {\bibfnamefont {J.~M.}\ \bibnamefont {Ok}}, \bibinfo
  {author} {\bibfnamefont {J.~S.}\ \bibnamefont {Kim}}, \bibinfo {author}
  {\bibfnamefont {J.}~\bibnamefont {van~den Brink}},\ and\ \bibinfo {author}
  {\bibfnamefont {B.}~\bibnamefont {Büchner}},\ }\bibfield  {title} {\bibinfo
  {title} {Nematicity and in-plane anisotropy of superconductivity in
  $\beta$-{FeSe} detected by $^{77}${Se} nuclear magnetic resonance},\ }\href
  {https://doi.org/10.1103/physrevb.93.180502} {\bibfield  {journal} {\bibinfo
  {journal} {Phys. Rev. B}\ }\textbf {\bibinfo {volume} {93}},\ \bibinfo
  {pages} {180502} (\bibinfo {year} {2016})}\BibitemShut {NoStop}%
\bibitem [{\citenamefont {Li}\ \emph {et~al.}(2020)\citenamefont {Li},
  \citenamefont {Lei}, \citenamefont {Zhao}, \citenamefont {Nie}, \citenamefont
  {Song}, \citenamefont {Zheng}, \citenamefont {Li}, \citenamefont {Kang},
  \citenamefont {Luo}, \citenamefont {Wu},\ and\ \citenamefont
  {Chen}}]{Li2020}%
  \BibitemOpen
  \bibfield  {author} {\bibinfo {author} {\bibfnamefont {J.}~\bibnamefont
  {Li}}, \bibinfo {author} {\bibfnamefont {B.}~\bibnamefont {Lei}}, \bibinfo
  {author} {\bibfnamefont {D.}~\bibnamefont {Zhao}}, \bibinfo {author}
  {\bibfnamefont {L.}~\bibnamefont {Nie}}, \bibinfo {author} {\bibfnamefont
  {D.}~\bibnamefont {Song}}, \bibinfo {author} {\bibfnamefont {L.}~\bibnamefont
  {Zheng}}, \bibinfo {author} {\bibfnamefont {S.}~\bibnamefont {Li}}, \bibinfo
  {author} {\bibfnamefont {B.}~\bibnamefont {Kang}}, \bibinfo {author}
  {\bibfnamefont {X.}~\bibnamefont {Luo}}, \bibinfo {author} {\bibfnamefont
  {T.}~\bibnamefont {Wu}},\ and\ \bibinfo {author} {\bibfnamefont
  {X.}~\bibnamefont {Chen}},\ }\bibfield  {title} {\bibinfo {title}
  {Spin-orbital-intertwined nematic state in {FeSe}},\ }\href
  {https://doi.org/10.1103/physrevx.10.011034} {\bibfield  {journal} {\bibinfo
  {journal} {Phys. Rev. X}\ }\textbf {\bibinfo {volume} {10}},\ \bibinfo
  {pages} {011034} (\bibinfo {year} {2020})}\BibitemShut {NoStop}%
\bibitem [{\citenamefont {{Molatta}}\ \emph {et~al.}(2020)\citenamefont
  {{Molatta}}, \citenamefont {{Opherden}}, \citenamefont {{Wosnitza}},
  \citenamefont {{Zhang}}, \citenamefont {{Wolf}}, \citenamefont
  {{L{\"o}hneysen}}, \citenamefont {{Sarkar}}, \citenamefont {{Biswas}},
  \citenamefont {{Grafe}},\ and\ \citenamefont {{K{\"u}hne}}}]{Molatta2020}%
  \BibitemOpen
  \bibfield  {author} {\bibinfo {author} {\bibfnamefont {S.}~\bibnamefont
  {{Molatta}}}, \bibinfo {author} {\bibfnamefont {D.}~\bibnamefont
  {{Opherden}}}, \bibinfo {author} {\bibfnamefont {J.}~\bibnamefont
  {{Wosnitza}}}, \bibinfo {author} {\bibfnamefont {Z.~T.}\ \bibnamefont
  {{Zhang}}}, \bibinfo {author} {\bibfnamefont {T.}~\bibnamefont {{Wolf}}},
  \bibinfo {author} {\bibfnamefont {H.~v.}\ \bibnamefont {{L{\"o}hneysen}}},
  \bibinfo {author} {\bibfnamefont {R.}~\bibnamefont {{Sarkar}}}, \bibinfo
  {author} {\bibfnamefont {P.~K.}\ \bibnamefont {{Biswas}}}, \bibinfo {author}
  {\bibfnamefont {H.~J.}\ \bibnamefont {{Grafe}}},\ and\ \bibinfo {author}
  {\bibfnamefont {H.}~\bibnamefont {{K{\"u}hne}}},\ }\bibfield  {title}
  {\bibinfo {title} {{Superconductivity of highly spin-polarized electrons in
  FeSe probed by $^{77}$Se NMR}},\ }\href@noop {} {\bibfield  {journal}
  {\bibinfo  {journal} {arXiv e-prints}\ ,\ \bibinfo {eid} {arXiv:2010.10128}}
  (\bibinfo {year} {2020})},\ \Eprint {https://arxiv.org/abs/2010.10128}
  {arXiv:2010.10128 [cond-mat.supr-con]} \BibitemShut {NoStop}%
\bibitem [{\citenamefont {Ishida}\ \emph {et~al.}(1998)\citenamefont {Ishida},
  \citenamefont {Mukuda}, \citenamefont {Kitaoka}, \citenamefont {Asayama},
  \citenamefont {Mao}, \citenamefont {Mori},\ and\ \citenamefont
  {Maeno}}]{IshidaSr2RuO4}%
  \BibitemOpen
  \bibfield  {author} {\bibinfo {author} {\bibfnamefont {K.}~\bibnamefont
  {Ishida}}, \bibinfo {author} {\bibfnamefont {H.}~\bibnamefont {Mukuda}},
  \bibinfo {author} {\bibfnamefont {Y.}~\bibnamefont {Kitaoka}}, \bibinfo
  {author} {\bibfnamefont {K.}~\bibnamefont {Asayama}}, \bibinfo {author}
  {\bibfnamefont {Z.~Q.}\ \bibnamefont {Mao}}, \bibinfo {author} {\bibfnamefont
  {Y.}~\bibnamefont {Mori}},\ and\ \bibinfo {author} {\bibfnamefont
  {Y.}~\bibnamefont {Maeno}},\ }\bibfield  {title} {\bibinfo {title}
  {Spin-triplet superconductivity in {Sr$_2$RuO$_4$} identified by {$^{17}$O}
  {K}night shift},\ }\href {https://doi.org/10.1038/25315} {\bibfield
  {journal} {\bibinfo  {journal} {Nature}\ }\textbf {\bibinfo {volume} {396}},\
  \bibinfo {pages} {658} (\bibinfo {year} {1998})}\BibitemShut {NoStop}%
\bibitem [{\citenamefont {Pustogow}\ \emph {et~al.}(2019)\citenamefont
  {Pustogow}, \citenamefont {Luo}, \citenamefont {Chronister}, \citenamefont
  {Su}, \citenamefont {Sokolov}, \citenamefont {Jerzembeck}, \citenamefont
  {Mackenzie}, \citenamefont {Hicks}, \citenamefont {Kikugawa}, \citenamefont
  {Raghu}, \citenamefont {Bauer},\ and\ \citenamefont {Brown}}]{Pustogow2019}%
  \BibitemOpen
  \bibfield  {author} {\bibinfo {author} {\bibfnamefont {A.}~\bibnamefont
  {Pustogow}}, \bibinfo {author} {\bibfnamefont {Y.}~\bibnamefont {Luo}},
  \bibinfo {author} {\bibfnamefont {A.}~\bibnamefont {Chronister}}, \bibinfo
  {author} {\bibfnamefont {Y.-S.}\ \bibnamefont {Su}}, \bibinfo {author}
  {\bibfnamefont {D.~A.}\ \bibnamefont {Sokolov}}, \bibinfo {author}
  {\bibfnamefont {F.}~\bibnamefont {Jerzembeck}}, \bibinfo {author}
  {\bibfnamefont {A.~P.}\ \bibnamefont {Mackenzie}}, \bibinfo {author}
  {\bibfnamefont {C.~W.}\ \bibnamefont {Hicks}}, \bibinfo {author}
  {\bibfnamefont {N.}~\bibnamefont {Kikugawa}}, \bibinfo {author}
  {\bibfnamefont {S.}~\bibnamefont {Raghu}}, \bibinfo {author} {\bibfnamefont
  {E.~D.}\ \bibnamefont {Bauer}},\ and\ \bibinfo {author} {\bibfnamefont
  {S.~E.}\ \bibnamefont {Brown}},\ }\bibfield  {title} {\bibinfo {title}
  {Constraints on the superconducting order parameter in {Sr$_2$RuO$_4$} from
  oxygen-17 nuclear magnetic resonance},\ }\href
  {https://doi.org/10.1038/s41586-019-1596-2} {\bibfield  {journal} {\bibinfo
  {journal} {Nature}\ }\textbf {\bibinfo {volume} {574}},\ \bibinfo {pages}
  {72} (\bibinfo {year} {2019})}\BibitemShut {NoStop}%
\bibitem [{\citenamefont {Cvetkovic}\ and\ \citenamefont
  {Vafek}(2013)}]{Vafek13}%
  \BibitemOpen
  \bibfield  {author} {\bibinfo {author} {\bibfnamefont {V.}~\bibnamefont
  {Cvetkovic}}\ and\ \bibinfo {author} {\bibfnamefont {O.}~\bibnamefont
  {Vafek}},\ }\bibfield  {title} {\bibinfo {title} {Space group symmetry,
  spin-orbit coupling, and the low-energy effective {H}amiltonian for
  iron-based superconductors},\ }\href
  {https://doi.org/10.1103/PhysRevB.88.134510} {\bibfield  {journal} {\bibinfo
  {journal} {Phys. Rev. B}\ }\textbf {\bibinfo {volume} {88}},\ \bibinfo
  {pages} {134510} (\bibinfo {year} {2013})}\BibitemShut {NoStop}%
\bibitem [{\citenamefont {Vafek}\ and\ \citenamefont
  {Chubukov}(2017)}]{Vafek2017}%
  \BibitemOpen
  \bibfield  {author} {\bibinfo {author} {\bibfnamefont {O.}~\bibnamefont
  {Vafek}}\ and\ \bibinfo {author} {\bibfnamefont {A.~V.}\ \bibnamefont
  {Chubukov}},\ }\bibfield  {title} {\bibinfo {title} {Hund interaction,
  spin-orbit coupling, and the mechanism of superconductivity in strongly
  hole-doped iron pnictides},\ }\href
  {https://doi.org/10.1103/physrevlett.118.087003} {\bibfield  {journal}
  {\bibinfo  {journal} {Phys. Rev. Lett.}\ }\textbf {\bibinfo {volume} {118}},\
  \bibinfo {pages} {087003} (\bibinfo {year} {2017})}\BibitemShut {NoStop}%
\bibitem [{\citenamefont {Böhmer}\ \emph {et~al.}(2016)\citenamefont
  {Böhmer}, \citenamefont {Taufour}, \citenamefont {Straszheim}, \citenamefont
  {Wolf},\ and\ \citenamefont {Canfield}}]{Boehmer2016}%
  \BibitemOpen
  \bibfield  {author} {\bibinfo {author} {\bibfnamefont {A.~E.}\ \bibnamefont
  {Böhmer}}, \bibinfo {author} {\bibfnamefont {V.}~\bibnamefont {Taufour}},
  \bibinfo {author} {\bibfnamefont {W.~E.}\ \bibnamefont {Straszheim}},
  \bibinfo {author} {\bibfnamefont {T.}~\bibnamefont {Wolf}},\ and\ \bibinfo
  {author} {\bibfnamefont {P.~C.}\ \bibnamefont {Canfield}},\ }\bibfield
  {title} {\bibinfo {title} {Variation of transition temperatures and residual
  resistivity ratio in vapor-grown {FeSe}},\ }\href
  {https://link.aps.org/doi/10.1103/PhysRevB.94.024526} {\bibfield  {journal}
  {\bibinfo  {journal} {Phys. Rev. B}\ }\textbf {\bibinfo {volume} {94}},\
  \bibinfo {pages} {024526} (\bibinfo {year} {2016})}\BibitemShut {NoStop}%
\bibitem [{\citenamefont {Vedeneev}\ \emph {et~al.}(2013)\citenamefont
  {Vedeneev}, \citenamefont {Piot}, \citenamefont {Maude},\ and\ \citenamefont
  {Sadakov}}]{Vedeneev2013}%
  \BibitemOpen
  \bibfield  {author} {\bibinfo {author} {\bibfnamefont {S.~I.}\ \bibnamefont
  {Vedeneev}}, \bibinfo {author} {\bibfnamefont {B.~A.}\ \bibnamefont {Piot}},
  \bibinfo {author} {\bibfnamefont {D.~K.}\ \bibnamefont {Maude}},\ and\
  \bibinfo {author} {\bibfnamefont {A.~V.}\ \bibnamefont {Sadakov}},\
  }\bibfield  {title} {\bibinfo {title} {Temperature dependence of the upper
  critical field of {FeSe} single crystals},\ }\href
  {https://doi.org/10.1103/physrevb.87.134512} {\bibfield  {journal} {\bibinfo
  {journal} {Phys. Rev. B}\ }\textbf {\bibinfo {volume} {87}},\ \bibinfo
  {pages} {134512} (\bibinfo {year} {2013})}\BibitemShut {NoStop}%
\bibitem [{\citenamefont {Zhou}\ \emph {et~al.}(2020)\citenamefont {Zhou},
  \citenamefont {Scherer}, \citenamefont {Mayaffre}, \citenamefont
  {Toulemonde}, \citenamefont {Ma}, \citenamefont {Li}, \citenamefont
  {Andersen},\ and\ \citenamefont {Julien}}]{Zhou2020}%
  \BibitemOpen
  \bibfield  {author} {\bibinfo {author} {\bibfnamefont {R.}~\bibnamefont
  {Zhou}}, \bibinfo {author} {\bibfnamefont {D.~D.}\ \bibnamefont {Scherer}},
  \bibinfo {author} {\bibfnamefont {H.}~\bibnamefont {Mayaffre}}, \bibinfo
  {author} {\bibfnamefont {P.}~\bibnamefont {Toulemonde}}, \bibinfo {author}
  {\bibfnamefont {M.}~\bibnamefont {Ma}}, \bibinfo {author} {\bibfnamefont
  {Y.}~\bibnamefont {Li}}, \bibinfo {author} {\bibfnamefont {B.~M.}\
  \bibnamefont {Andersen}},\ and\ \bibinfo {author} {\bibfnamefont {M.-H.}\
  \bibnamefont {Julien}},\ }\bibfield  {title} {\bibinfo {title} {Singular
  magnetic anisotropy in the nematic phase of {FeSe}},\ }\href
  {https://doi.org/10.1038/s41535-020-00295-1} {\bibfield  {journal} {\bibinfo
  {journal} {npj Quantum Materials}\ }\textbf {\bibinfo {volume} {5}},\
  \bibinfo {pages} {93} (\bibinfo {year} {2020})}\BibitemShut {NoStop}%
\bibitem [{\citenamefont {Cao}\ \emph {et~al.}(2019)\citenamefont {Cao},
  \citenamefont {Hu}, \citenamefont {Dong}, \citenamefont {Zhang},
  \citenamefont {Ye}, \citenamefont {Xu}, \citenamefont {Chareev},
  \citenamefont {Vasiliev}, \citenamefont {Wu}, \citenamefont {Zeng},
  \citenamefont {Wang},\ and\ \citenamefont {Wu}}]{Cao2019}%
  \BibitemOpen
  \bibfield  {author} {\bibinfo {author} {\bibfnamefont {R.~X.}\ \bibnamefont
  {Cao}}, \bibinfo {author} {\bibfnamefont {J.}~\bibnamefont {Hu}}, \bibinfo
  {author} {\bibfnamefont {J.}~\bibnamefont {Dong}}, \bibinfo {author}
  {\bibfnamefont {J.~B.}\ \bibnamefont {Zhang}}, \bibinfo {author}
  {\bibfnamefont {X.~S.}\ \bibnamefont {Ye}}, \bibinfo {author} {\bibfnamefont
  {Y.~F.}\ \bibnamefont {Xu}}, \bibinfo {author} {\bibfnamefont {D.~A.}\
  \bibnamefont {Chareev}}, \bibinfo {author} {\bibfnamefont {A.~N.}\
  \bibnamefont {Vasiliev}}, \bibinfo {author} {\bibfnamefont {B.}~\bibnamefont
  {Wu}}, \bibinfo {author} {\bibfnamefont {X.~H.}\ \bibnamefont {Zeng}},
  \bibinfo {author} {\bibfnamefont {Q.~L.}\ \bibnamefont {Wang}},\ and\
  \bibinfo {author} {\bibfnamefont {G.}~\bibnamefont {Wu}},\ }\bibfield
  {title} {\bibinfo {title} {Observation of orbital ordering and origin of the
  nematic order in {FeSe}},\ }\href {https://doi.org/10.1088/1367-2630/ab4927}
  {\bibfield  {journal} {\bibinfo  {journal} {New J. Phys.}\ }\textbf {\bibinfo
  {volume} {21}},\ \bibinfo {pages} {103033} (\bibinfo {year}
  {2019})}\BibitemShut {NoStop}%
\bibitem [{\citenamefont {Shirer}\ \emph {et~al.}(2012)\citenamefont {Shirer},
  \citenamefont {Shockley}, \citenamefont {Dioguardi}, \citenamefont {Crocker},
  \citenamefont {Lin}, \citenamefont {apRoberts Warren}, \citenamefont
  {Nisson}, \citenamefont {Klavins}, \citenamefont {Cooley}, \citenamefont
  {Yang},\ and\ \citenamefont {Curro}}]{ShirerPNAS2012}%
  \BibitemOpen
  \bibfield  {author} {\bibinfo {author} {\bibfnamefont {K.~R.}\ \bibnamefont
  {Shirer}}, \bibinfo {author} {\bibfnamefont {A.~C.}\ \bibnamefont
  {Shockley}}, \bibinfo {author} {\bibfnamefont {A.~P.}\ \bibnamefont
  {Dioguardi}}, \bibinfo {author} {\bibfnamefont {J.}~\bibnamefont {Crocker}},
  \bibinfo {author} {\bibfnamefont {C.~H.}\ \bibnamefont {Lin}}, \bibinfo
  {author} {\bibfnamefont {N.}~\bibnamefont {apRoberts Warren}}, \bibinfo
  {author} {\bibfnamefont {D.~M.}\ \bibnamefont {Nisson}}, \bibinfo {author}
  {\bibfnamefont {P.}~\bibnamefont {Klavins}}, \bibinfo {author} {\bibfnamefont
  {J.~C.}\ \bibnamefont {Cooley}}, \bibinfo {author} {\bibfnamefont {Y.-f.}\
  \bibnamefont {Yang}},\ and\ \bibinfo {author} {\bibfnamefont {N.~J.}\
  \bibnamefont {Curro}},\ }\bibfield  {title} {\bibinfo {title} {Long range
  order and two-fluid behavior in heavy electron materials},\ }\href
  {https://doi.org/10.1073/pnas.1209609109} {\bibfield  {journal} {\bibinfo
  {journal} {Proc. Natl. Acad. Sci.}\ }\textbf {\bibinfo {volume} {109}},\
  \bibinfo {pages} {E3067} (\bibinfo {year} {2012})}\BibitemShut {NoStop}%
\bibitem [{\citenamefont {Nisson}\ and\ \citenamefont
  {Curro}(2016)}]{NissonCEFSOC2016}%
  \BibitemOpen
  \bibfield  {author} {\bibinfo {author} {\bibfnamefont {D.~M.}\ \bibnamefont
  {Nisson}}\ and\ \bibinfo {author} {\bibfnamefont {N.~J.}\ \bibnamefont
  {Curro}},\ }\bibfield  {title} {\bibinfo {title} {Nuclear magnetic resonance
  {K}night shifts in the presence of strong spin-orbit and crystal-field
  potentials},\ }\href {http://stacks.iop.org/1367-2630/18/i=7/a=073041}
  {\bibfield  {journal} {\bibinfo  {journal} {New J. Phys.}\ }\textbf {\bibinfo
  {volume} {18}},\ \bibinfo {pages} {073041} (\bibinfo {year}
  {2016})}\BibitemShut {NoStop}%
\bibitem [{\citenamefont {Kotegawa}\ \emph {et~al.}(2008)\citenamefont
  {Kotegawa}, \citenamefont {Masaki}, \citenamefont {Awai}, \citenamefont
  {Tou}, \citenamefont {Mizuguchi},\ and\ \citenamefont
  {Takano}}]{Kotegawa2008}%
  \BibitemOpen
  \bibfield  {author} {\bibinfo {author} {\bibfnamefont {H.}~\bibnamefont
  {Kotegawa}}, \bibinfo {author} {\bibfnamefont {S.}~\bibnamefont {Masaki}},
  \bibinfo {author} {\bibfnamefont {Y.}~\bibnamefont {Awai}}, \bibinfo {author}
  {\bibfnamefont {H.}~\bibnamefont {Tou}}, \bibinfo {author} {\bibfnamefont
  {Y.}~\bibnamefont {Mizuguchi}},\ and\ \bibinfo {author} {\bibfnamefont
  {Y.}~\bibnamefont {Takano}},\ }\bibfield  {title} {\bibinfo {title} {Evidence
  for unconventional superconductivity in arsenic-free iron-based
  superconductor {FeSe}: A {$^{77}$Se-NMR} study},\ }\href
  {https://doi.org/10.1143/jpsj.77.113703} {\bibfield  {journal} {\bibinfo
  {journal} {J. Phys. Soc. Japan}\ }\textbf {\bibinfo {volume} {77}},\ \bibinfo
  {pages} {113703} (\bibinfo {year} {2008})}\BibitemShut {NoStop}%
\bibitem [{\citenamefont {Curro}\ \emph {et~al.}(2005)\citenamefont {Curro},
  \citenamefont {Caldwell}, \citenamefont {Bauer}, \citenamefont {Morales},
  \citenamefont {Graf}, \citenamefont {Bang}, \citenamefont {Balatsky},
  \citenamefont {Thompson},\ and\ \citenamefont {Sarrao}}]{Curro2005}%
  \BibitemOpen
  \bibfield  {author} {\bibinfo {author} {\bibfnamefont {N.}~\bibnamefont
  {Curro}}, \bibinfo {author} {\bibfnamefont {T.}~\bibnamefont {Caldwell}},
  \bibinfo {author} {\bibfnamefont {E.}~\bibnamefont {Bauer}}, \bibinfo
  {author} {\bibfnamefont {L.}~\bibnamefont {Morales}}, \bibinfo {author}
  {\bibfnamefont {M.}~\bibnamefont {Graf}}, \bibinfo {author} {\bibfnamefont
  {Y.}~\bibnamefont {Bang}}, \bibinfo {author} {\bibfnamefont {A.}~\bibnamefont
  {Balatsky}}, \bibinfo {author} {\bibfnamefont {J.}~\bibnamefont {Thompson}},\
  and\ \bibinfo {author} {\bibfnamefont {J.}~\bibnamefont {Sarrao}},\
  }\bibfield  {title} {\bibinfo {title} {Unconventional superconductivity in
  {PuCoGa$_5$}},\ }\href {https://doi.org/10.1038/nature03428} {\bibfield
  {journal} {\bibinfo  {journal} {Nature}\ }\textbf {\bibinfo {volume} {434}},\
  \bibinfo {pages} {622} (\bibinfo {year} {2005})}\BibitemShut {NoStop}%
\bibitem [{\citenamefont {Tan}\ \emph {et~al.}(2013)\citenamefont {Tan},
  \citenamefont {Zhang}, \citenamefont {Xia}, \citenamefont {Ye}, \citenamefont
  {Chen}, \citenamefont {Xie}, \citenamefont {Peng}, \citenamefont {Xu},
  \citenamefont {Fan}, \citenamefont {Xu}, \citenamefont {Jiang}, \citenamefont
  {Zhang}, \citenamefont {Lai}, \citenamefont {Xiang}, \citenamefont {Hu},
  \citenamefont {Xie},\ and\ \citenamefont {Feng}}]{Tan2013}%
  \BibitemOpen
  \bibfield  {author} {\bibinfo {author} {\bibfnamefont {S.}~\bibnamefont
  {Tan}}, \bibinfo {author} {\bibfnamefont {Y.}~\bibnamefont {Zhang}}, \bibinfo
  {author} {\bibfnamefont {M.}~\bibnamefont {Xia}}, \bibinfo {author}
  {\bibfnamefont {Z.}~\bibnamefont {Ye}}, \bibinfo {author} {\bibfnamefont
  {F.}~\bibnamefont {Chen}}, \bibinfo {author} {\bibfnamefont {X.}~\bibnamefont
  {Xie}}, \bibinfo {author} {\bibfnamefont {R.}~\bibnamefont {Peng}}, \bibinfo
  {author} {\bibfnamefont {D.}~\bibnamefont {Xu}}, \bibinfo {author}
  {\bibfnamefont {Q.}~\bibnamefont {Fan}}, \bibinfo {author} {\bibfnamefont
  {H.}~\bibnamefont {Xu}}, \bibinfo {author} {\bibfnamefont {J.}~\bibnamefont
  {Jiang}}, \bibinfo {author} {\bibfnamefont {T.}~\bibnamefont {Zhang}},
  \bibinfo {author} {\bibfnamefont {X.}~\bibnamefont {Lai}}, \bibinfo {author}
  {\bibfnamefont {T.}~\bibnamefont {Xiang}}, \bibinfo {author} {\bibfnamefont
  {J.}~\bibnamefont {Hu}}, \bibinfo {author} {\bibfnamefont {B.}~\bibnamefont
  {Xie}},\ and\ \bibinfo {author} {\bibfnamefont {D.}~\bibnamefont {Feng}},\
  }\bibfield  {title} {\bibinfo {title} {Interface-induced superconductivity
  and strain-dependent spin density waves in {FeSe}/{SrTiO}3~thin films},\
  }\href {https://doi.org/10.1038/nmat3654} {\bibfield  {journal} {\bibinfo
  {journal} {Nat. Mater.}\ }\textbf {\bibinfo {volume} {12}},\ \bibinfo {pages}
  {634} (\bibinfo {year} {2013})}\BibitemShut {NoStop}%
\bibitem [{\citenamefont {Sonier}(2007)}]{Sonier2007}%
  \BibitemOpen
  \bibfield  {author} {\bibinfo {author} {\bibfnamefont {J.~E.}\ \bibnamefont
  {Sonier}},\ }\bibfield  {title} {\bibinfo {title} {Muon spin rotation studies
  of electronic excitations and magnetism in the vortex cores of
  superconductors},\ }\href {https://doi.org/10.1088/0034-4885/70/11/r01}
  {\bibfield  {journal} {\bibinfo  {journal} {Rep. Progr. Phys.}\ }\textbf
  {\bibinfo {volume} {70}},\ \bibinfo {pages} {1717} (\bibinfo {year}
  {2007})}\BibitemShut {NoStop}%
\bibitem [{\citenamefont {Abdel-Hafiez}\ \emph {et~al.}(2013)\citenamefont
  {Abdel-Hafiez}, \citenamefont {Ge}, \citenamefont {Vasiliev}, \citenamefont
  {Chareev}, \citenamefont {de~Vondel}, \citenamefont {Moshchalkov},\ and\
  \citenamefont {Silhanek}}]{AbdelHafiez2013}%
  \BibitemOpen
  \bibfield  {author} {\bibinfo {author} {\bibfnamefont {M.}~\bibnamefont
  {Abdel-Hafiez}}, \bibinfo {author} {\bibfnamefont {J.}~\bibnamefont {Ge}},
  \bibinfo {author} {\bibfnamefont {A.~N.}\ \bibnamefont {Vasiliev}}, \bibinfo
  {author} {\bibfnamefont {D.~A.}\ \bibnamefont {Chareev}}, \bibinfo {author}
  {\bibfnamefont {J.~V.}\ \bibnamefont {de~Vondel}}, \bibinfo {author}
  {\bibfnamefont {V.~V.}\ \bibnamefont {Moshchalkov}},\ and\ \bibinfo {author}
  {\bibfnamefont {A.~V.}\ \bibnamefont {Silhanek}},\ }\bibfield  {title}
  {\bibinfo {title} {Temperature dependence of lower critical field
  {H$_{c1}$(T)} shows nodeless superconductivity in {FeSe}},\ }\href
  {https://doi.org/10.1103/physrevb.88.174512} {\bibfield  {journal} {\bibinfo
  {journal} {Phys. Rev. B}\ }\textbf {\bibinfo {volume} {88}},\ \bibinfo
  {pages} {174512} (\bibinfo {year} {2013})}\BibitemShut {NoStop}%
\bibitem [{\citenamefont {Maisuradze}\ \emph {et~al.}(2009)\citenamefont
  {Maisuradze}, \citenamefont {Khasanov}, \citenamefont {Shengelaya},\ and\
  \citenamefont {Keller}}]{Maisuradze2009}%
  \BibitemOpen
  \bibfield  {author} {\bibinfo {author} {\bibfnamefont {A.}~\bibnamefont
  {Maisuradze}}, \bibinfo {author} {\bibfnamefont {R.}~\bibnamefont
  {Khasanov}}, \bibinfo {author} {\bibfnamefont {A.}~\bibnamefont
  {Shengelaya}},\ and\ \bibinfo {author} {\bibfnamefont {H.}~\bibnamefont
  {Keller}},\ }\bibfield  {title} {\bibinfo {title} {Comparison of different
  methods for analyzing $\mu${SR} line shapes in the vortex state of type-{II}
  superconductors},\ }\href {https://doi.org/10.1088/0953-8984/21/7/075701}
  {\bibfield  {journal} {\bibinfo  {journal} {J. Phys.: Condens. Matter}\
  }\textbf {\bibinfo {volume} {21}},\ \bibinfo {pages} {075701} (\bibinfo
  {year} {2009})}\BibitemShut {NoStop}%
\bibitem [{sup()}]{supplemental}%
  \BibitemOpen
  \href@noop {} {}\bibinfo {note} {See Supplemental Material at [URL will be
  inserted by publisher] for [give brief description of material]}\BibitemShut
  {NoStop}%
\bibitem [{\citenamefont {Shi}\ \emph {et~al.}(2018)\citenamefont {Shi},
  \citenamefont {Arai}, \citenamefont {Kitagawa}, \citenamefont {Yamanaka},
  \citenamefont {Ishida}, \citenamefont {Böhmer}, \citenamefont {Meingast},
  \citenamefont {Wolf}, \citenamefont {Hirata},\ and\ \citenamefont
  {Sasaki}}]{Shi2018}%
  \BibitemOpen
  \bibfield  {author} {\bibinfo {author} {\bibfnamefont {A.}~\bibnamefont
  {Shi}}, \bibinfo {author} {\bibfnamefont {T.}~\bibnamefont {Arai}}, \bibinfo
  {author} {\bibfnamefont {S.}~\bibnamefont {Kitagawa}}, \bibinfo {author}
  {\bibfnamefont {T.}~\bibnamefont {Yamanaka}}, \bibinfo {author}
  {\bibfnamefont {K.}~\bibnamefont {Ishida}}, \bibinfo {author} {\bibfnamefont
  {A.~E.}\ \bibnamefont {Böhmer}}, \bibinfo {author} {\bibfnamefont
  {C.}~\bibnamefont {Meingast}}, \bibinfo {author} {\bibfnamefont
  {T.}~\bibnamefont {Wolf}}, \bibinfo {author} {\bibfnamefont {M.}~\bibnamefont
  {Hirata}},\ and\ \bibinfo {author} {\bibfnamefont {T.}~\bibnamefont
  {Sasaki}},\ }\bibfield  {title} {\bibinfo {title} {Pseudogap behavior of the
  nuclear spin{\textendash}lattice relaxation rate in {FeSe} probed by
  $^{77}${Se-NMR}},\ }\href {https://doi.org/10.7566/jpsj.87.013704} {\bibfield
   {journal} {\bibinfo  {journal} {J. Phys. Soc. Japan}\ }\textbf {\bibinfo
  {volume} {87}},\ \bibinfo {pages} {013704} (\bibinfo {year}
  {2018})}\BibitemShut {NoStop}%
\bibitem [{\citenamefont {Caroli}\ \emph {et~al.}(1964)\citenamefont {Caroli},
  \citenamefont {Gennes},\ and\ \citenamefont {Matricon}}]{Caroli1964}%
  \BibitemOpen
  \bibfield  {author} {\bibinfo {author} {\bibfnamefont {C.}~\bibnamefont
  {Caroli}}, \bibinfo {author} {\bibfnamefont {P.~D.}\ \bibnamefont {Gennes}},\
  and\ \bibinfo {author} {\bibfnamefont {J.}~\bibnamefont {Matricon}},\
  }\bibfield  {title} {\bibinfo {title} {Bound fermion states on a vortex line
  in a type {II} superconductor},\ }\href
  {https://doi.org/10.1016/0031-9163(64)90375-0} {\bibfield  {journal}
  {\bibinfo  {journal} {Phys. Lett. A}\ }\textbf {\bibinfo {volume} {9}},\
  \bibinfo {pages} {307} (\bibinfo {year} {1964})}\BibitemShut {NoStop}%
\bibitem [{\citenamefont {Dukan}\ and\ \citenamefont
  {Te{\v{s}}anovi{\'{c}}}(1994)}]{Dukan1994}%
  \BibitemOpen
  \bibfield  {author} {\bibinfo {author} {\bibfnamefont {S.}~\bibnamefont
  {Dukan}}\ and\ \bibinfo {author} {\bibfnamefont {Z.}~\bibnamefont
  {Te{\v{s}}anovi{\'{c}}}},\ }\bibfield  {title} {\bibinfo {title}
  {Superconductivity in a high magnetic field: Excitation spectrum and
  tunneling properties},\ }\href {https://doi.org/10.1103/physrevb.49.13017}
  {\bibfield  {journal} {\bibinfo  {journal} {Phys. Rev. B}\ }\textbf {\bibinfo
  {volume} {49}},\ \bibinfo {pages} {13017} (\bibinfo {year}
  {1994})}\BibitemShut {NoStop}%
\bibitem [{\citenamefont {Norman}\ \emph {et~al.}(1995)\citenamefont {Norman},
  \citenamefont {MacDonald},\ and\ \citenamefont {Akera}}]{Norman1995}%
  \BibitemOpen
  \bibfield  {author} {\bibinfo {author} {\bibfnamefont {M.~R.}\ \bibnamefont
  {Norman}}, \bibinfo {author} {\bibfnamefont {A.~H.}\ \bibnamefont
  {MacDonald}},\ and\ \bibinfo {author} {\bibfnamefont {H.}~\bibnamefont
  {Akera}},\ }\bibfield  {title} {\bibinfo {title} {Magnetic oscillations and
  quasiparticle band structure in the mixed state of type-{II}
  superconductors},\ }\href {https://doi.org/10.1103/physrevb.51.5927}
  {\bibfield  {journal} {\bibinfo  {journal} {Phys. Rev. B}\ }\textbf {\bibinfo
  {volume} {51}},\ \bibinfo {pages} {5927} (\bibinfo {year}
  {1995})}\BibitemShut {NoStop}%
\bibitem [{\citenamefont {Ichioka}\ \emph {et~al.}(1999)\citenamefont
  {Ichioka}, \citenamefont {Hasegawa},\ and\ \citenamefont
  {Machida}}]{Ichioka1999}%
  \BibitemOpen
  \bibfield  {author} {\bibinfo {author} {\bibfnamefont {M.}~\bibnamefont
  {Ichioka}}, \bibinfo {author} {\bibfnamefont {A.}~\bibnamefont {Hasegawa}},\
  and\ \bibinfo {author} {\bibfnamefont {K.}~\bibnamefont {Machida}},\
  }\bibfield  {title} {\bibinfo {title} {Field dependence of the vortex
  structure ind-wave ands-wave superconductors},\ }\href
  {https://doi.org/10.1103/physrevb.59.8902} {\bibfield  {journal} {\bibinfo
  {journal} {Phys. Rev. B}\ }\textbf {\bibinfo {volume} {59}},\ \bibinfo
  {pages} {8902} (\bibinfo {year} {1999})}\BibitemShut {NoStop}%
\bibitem [{\citenamefont {Zhu}(2016)}]{Zhu2016}%
  \BibitemOpen
  \bibfield  {author} {\bibinfo {author} {\bibfnamefont {J.-X.}\ \bibnamefont
  {Zhu}},\ }\bibinfo {title} {Local electronic structure in superconductors
  under a magnetic field},\ in\ \href
  {https://doi.org/10.1007/978-3-319-31314-6_5} {\emph {\bibinfo {booktitle}
  {Bogoliubov-de Gennes Method and Its Applications}}}\ (\bibinfo  {publisher}
  {Springer International Publishing},\ \bibinfo {address} {Cham},\ \bibinfo
  {year} {2016})\ pp.\ \bibinfo {pages} {111--139}\BibitemShut {NoStop}%
\bibitem [{\citenamefont {Nakai}\ \emph {et~al.}(2008)\citenamefont {Nakai},
  \citenamefont {Hayashi}, \citenamefont {Kitagawa}, \citenamefont {Ishida},
  \citenamefont {Sugawara}, \citenamefont {Kikuchi},\ and\ \citenamefont
  {Sato}}]{Nakai2008a}%
  \BibitemOpen
  \bibfield  {author} {\bibinfo {author} {\bibfnamefont {Y.}~\bibnamefont
  {Nakai}}, \bibinfo {author} {\bibfnamefont {Y.}~\bibnamefont {Hayashi}},
  \bibinfo {author} {\bibfnamefont {K.}~\bibnamefont {Kitagawa}}, \bibinfo
  {author} {\bibfnamefont {K.}~\bibnamefont {Ishida}}, \bibinfo {author}
  {\bibfnamefont {H.}~\bibnamefont {Sugawara}}, \bibinfo {author}
  {\bibfnamefont {D.}~\bibnamefont {Kikuchi}},\ and\ \bibinfo {author}
  {\bibfnamefont {H.}~\bibnamefont {Sato}},\ }\bibfield  {title} {\bibinfo
  {title} {Evidence of the bound states of the vortex state in ans-wave
  superconductor proved by {NMR} measurements},\ }\href
  {https://doi.org/10.1143/jpsjs.77sa.333} {\bibfield  {journal} {\bibinfo
  {journal} {J. Phys. Soc. Japan}\ }\textbf {\bibinfo {volume} {77}},\ \bibinfo
  {pages} {333} (\bibinfo {year} {2008})}\BibitemShut {NoStop}%
\bibitem [{\citenamefont {Mitrovi\ifmmode~\acute{c}\else \'{c}\fi{}}\ \emph
  {et~al.}(2003)\citenamefont {Mitrovi\ifmmode~\acute{c}\else \'{c}\fi{}},
  \citenamefont {Sigmund}, \citenamefont {Halperin}, \citenamefont {Reyes},
  \citenamefont {Kuhns},\ and\ \citenamefont
  {Moulton}}]{AFMinVortexMitrovicPRB2003}%
  \BibitemOpen
  \bibfield  {author} {\bibinfo {author} {\bibfnamefont {V.~F.}\ \bibnamefont
  {Mitrovi\ifmmode~\acute{c}\else \'{c}\fi{}}}, \bibinfo {author}
  {\bibfnamefont {E.~E.}\ \bibnamefont {Sigmund}}, \bibinfo {author}
  {\bibfnamefont {W.~P.}\ \bibnamefont {Halperin}}, \bibinfo {author}
  {\bibfnamefont {A.~P.}\ \bibnamefont {Reyes}}, \bibinfo {author}
  {\bibfnamefont {P.}~\bibnamefont {Kuhns}},\ and\ \bibinfo {author}
  {\bibfnamefont {W.~G.}\ \bibnamefont {Moulton}},\ }\bibfield  {title}
  {\bibinfo {title} {Antiferromagnetism in the vortex cores of
  {YBa$_2$Cu$_3$O$_{7-\delta}$}},\ }\href
  {https://doi.org/10.1103/PhysRevB.67.220503} {\bibfield  {journal} {\bibinfo
  {journal} {Phys. Rev. B}\ }\textbf {\bibinfo {volume} {67}},\ \bibinfo
  {pages} {220503} (\bibinfo {year} {2003})}\BibitemShut {NoStop}%
\bibitem [{\citenamefont {Mitrovic}\ \emph {et~al.}(2001)\citenamefont
  {Mitrovic}, \citenamefont {Sigmund}, \citenamefont {Eschrig}, \citenamefont
  {Bachman}, \citenamefont {Halperin}, \citenamefont {Reyes}, \citenamefont
  {Kuhns},\ and\ \citenamefont {Moulton}}]{mitrovicYBCO}%
  \BibitemOpen
  \bibfield  {author} {\bibinfo {author} {\bibfnamefont {V.~F.}\ \bibnamefont
  {Mitrovic}}, \bibinfo {author} {\bibfnamefont {E.~E.}\ \bibnamefont
  {Sigmund}}, \bibinfo {author} {\bibfnamefont {M.}~\bibnamefont {Eschrig}},
  \bibinfo {author} {\bibfnamefont {H.~N.}\ \bibnamefont {Bachman}}, \bibinfo
  {author} {\bibfnamefont {W.~P.}\ \bibnamefont {Halperin}}, \bibinfo {author}
  {\bibfnamefont {A.~P.}\ \bibnamefont {Reyes}}, \bibinfo {author}
  {\bibfnamefont {P.}~\bibnamefont {Kuhns}},\ and\ \bibinfo {author}
  {\bibfnamefont {W.~G.}\ \bibnamefont {Moulton}},\ }\bibfield  {title}
  {\bibinfo {title} {Spatially resolved electronic structure inside and outside
  the vortex cores of a high-temperature superconductor},\ }\href
  {https://doi.org/10.1038/35097039} {\bibfield  {journal} {\bibinfo  {journal}
  {Nature}\ }\textbf {\bibinfo {volume} {413}},\ \bibinfo {pages} {6855}
  (\bibinfo {year} {2001})}\BibitemShut {NoStop}%
\bibitem [{\citenamefont {Curro}\ \emph {et~al.}(2000)\citenamefont {Curro},
  \citenamefont {Milling}, \citenamefont {Haase},\ and\ \citenamefont
  {Slichter}}]{Curro2000c}%
  \BibitemOpen
  \bibfield  {author} {\bibinfo {author} {\bibfnamefont {N.}~\bibnamefont
  {Curro}}, \bibinfo {author} {\bibfnamefont {C.}~\bibnamefont {Milling}},
  \bibinfo {author} {\bibfnamefont {J.}~\bibnamefont {Haase}},\ and\ \bibinfo
  {author} {\bibfnamefont {C.}~\bibnamefont {Slichter}},\ }\bibfield  {title}
  {\bibinfo {title} {Local-field dependence of the {O-17} spin-lattice
  relaxation and echo decay rates in the mixed state of {YBa$_2$Cu$_3$O$_7$}},\
  }\href {https://doi.org/10.1103/PhysRevB.62.3473} {\bibfield  {journal}
  {\bibinfo  {journal} {Phys. Rev. B}\ }\textbf {\bibinfo {volume} {62}},\
  \bibinfo {pages} {3473} (\bibinfo {year} {2000})}\BibitemShut {NoStop}%
\bibitem [{\citenamefont {Imai}\ \emph {et~al.}(2009)\citenamefont {Imai},
  \citenamefont {Ahilan}, \citenamefont {Ning}, \citenamefont {McQueen},\ and\
  \citenamefont {Cava}}]{Imai2009}%
  \BibitemOpen
  \bibfield  {author} {\bibinfo {author} {\bibfnamefont {T.}~\bibnamefont
  {Imai}}, \bibinfo {author} {\bibfnamefont {K.}~\bibnamefont {Ahilan}},
  \bibinfo {author} {\bibfnamefont {F.~L.}\ \bibnamefont {Ning}}, \bibinfo
  {author} {\bibfnamefont {T.~M.}\ \bibnamefont {McQueen}},\ and\ \bibinfo
  {author} {\bibfnamefont {R.~J.}\ \bibnamefont {Cava}},\ }\bibfield  {title}
  {\bibinfo {title} {Why does undoped {FeSe} become a high-{Tc} superconductor
  under pressure?},\ }\href {https://doi.org/10.1103/physrevlett.102.177005}
  {\bibfield  {journal} {\bibinfo  {journal} {Phys. Rev. Lett.}\ }\textbf
  {\bibinfo {volume} {102}},\ \bibinfo {pages} {177005} (\bibinfo {year}
  {2009})}\BibitemShut {NoStop}%
\bibitem [{\citenamefont {Mukherjee}\ \emph {et~al.}(2015)\citenamefont
  {Mukherjee}, \citenamefont {Kreisel}, \citenamefont {Hirschfeld},\ and\
  \citenamefont {Andersen}}]{Mukherjee2015}%
  \BibitemOpen
  \bibfield  {author} {\bibinfo {author} {\bibfnamefont {S.}~\bibnamefont
  {Mukherjee}}, \bibinfo {author} {\bibfnamefont {A.}~\bibnamefont {Kreisel}},
  \bibinfo {author} {\bibfnamefont {P.}~\bibnamefont {Hirschfeld}},\ and\
  \bibinfo {author} {\bibfnamefont {B.~M.}\ \bibnamefont {Andersen}},\
  }\bibfield  {title} {\bibinfo {title} {Model of electronic structure and
  superconductivity in orbitally ordered {FeSe}},\ }\href
  {https://doi.org/10.1103/physrevlett.115.026402} {\bibfield  {journal}
  {\bibinfo  {journal} {Phys. Rev. Lett.}\ }\textbf {\bibinfo {volume} {115}},\
  \bibinfo {pages} {026402} (\bibinfo {year} {2015})}\BibitemShut {NoStop}%
\bibitem [{\citenamefont {{Zhou}}\ \emph {et~al.}(2021)\citenamefont {{Zhou}},
  \citenamefont {{Sun}}, \citenamefont {{Xi}}, \citenamefont {{Wang}},
  \citenamefont {{Zhang}}, \citenamefont {{Xu}}, \citenamefont {{Pan}},
  \citenamefont {{Feng}}, \citenamefont {{Meng}}, \citenamefont {{Yi}},
  \citenamefont {{Pi}}, \citenamefont {{Tamegai}}, \citenamefont {{Xing}},\
  and\ \citenamefont {{Shi}}}]{ZhouFeSeDisorder2021}%
  \BibitemOpen
  \bibfield  {author} {\bibinfo {author} {\bibfnamefont {N.}~\bibnamefont
  {{Zhou}}}, \bibinfo {author} {\bibfnamefont {Y.}~\bibnamefont {{Sun}}},
  \bibinfo {author} {\bibfnamefont {C.~Y.}\ \bibnamefont {{Xi}}}, \bibinfo
  {author} {\bibfnamefont {Z.~S.}\ \bibnamefont {{Wang}}}, \bibinfo {author}
  {\bibfnamefont {Y.~F.}\ \bibnamefont {{Zhang}}}, \bibinfo {author}
  {\bibfnamefont {C.~Q.}\ \bibnamefont {{Xu}}}, \bibinfo {author}
  {\bibfnamefont {Y.~Q.}\ \bibnamefont {{Pan}}}, \bibinfo {author}
  {\bibfnamefont {J.~J.}\ \bibnamefont {{Feng}}}, \bibinfo {author}
  {\bibfnamefont {Y.}~\bibnamefont {{Meng}}}, \bibinfo {author} {\bibfnamefont
  {X.~L.}\ \bibnamefont {{Yi}}}, \bibinfo {author} {\bibfnamefont
  {L.}~\bibnamefont {{Pi}}}, \bibinfo {author} {\bibfnamefont {T.}~\bibnamefont
  {{Tamegai}}}, \bibinfo {author} {\bibfnamefont {X.}~\bibnamefont {{Xing}}},\
  and\ \bibinfo {author} {\bibfnamefont {Z.}~\bibnamefont {{Shi}}},\ }\bibfield
   {title} {\bibinfo {title} {{Disorder-robust high-field superconducting phase
  of FeSe single crystals}},\ }\href@noop {} {\bibfield  {journal} {\bibinfo
  {journal} {arXiv e-prints}\ ,\ \bibinfo {eid} {arXiv:2102.02353}} (\bibinfo
  {year} {2021})},\ \Eprint {https://arxiv.org/abs/2102.02353}
  {arXiv:2102.02353 [cond-mat.supr-con]} \BibitemShut {NoStop}%
\bibitem [{\citenamefont {Larsen}\ \emph {et~al.}(2015)\citenamefont {Larsen},
  \citenamefont {Uranga}, \citenamefont {Stieper}, \citenamefont {Holm},
  \citenamefont {Bernhard}, \citenamefont {Wolf}, \citenamefont {Lefmann},
  \citenamefont {Andersen},\ and\ \citenamefont {Niedermayer}}]{Larsen2015}%
  \BibitemOpen
  \bibfield  {author} {\bibinfo {author} {\bibfnamefont {J.}~\bibnamefont
  {Larsen}}, \bibinfo {author} {\bibfnamefont {B.~M.}\ \bibnamefont {Uranga}},
  \bibinfo {author} {\bibfnamefont {G.}~\bibnamefont {Stieper}}, \bibinfo
  {author} {\bibfnamefont {S.~L.}\ \bibnamefont {Holm}}, \bibinfo {author}
  {\bibfnamefont {C.}~\bibnamefont {Bernhard}}, \bibinfo {author}
  {\bibfnamefont {T.}~\bibnamefont {Wolf}}, \bibinfo {author} {\bibfnamefont
  {K.}~\bibnamefont {Lefmann}}, \bibinfo {author} {\bibfnamefont {B.~M.}\
  \bibnamefont {Andersen}},\ and\ \bibinfo {author} {\bibfnamefont
  {C.}~\bibnamefont {Niedermayer}},\ }\bibfield  {title} {\bibinfo {title}
  {Competing superconducting and magnetic order parameters and field-induced
  magnetism in electron-{dopedBa}(fe1-{xCox})2as2},\ }\href
  {https://doi.org/10.1103/physrevb.91.024504} {\bibfield  {journal} {\bibinfo
  {journal} {Physical Review B}\ }\textbf {\bibinfo {volume} {91}},\ \bibinfo
  {pages} {024504} (\bibinfo {year} {2015})}\BibitemShut {NoStop}%
\bibitem [{\citenamefont {Young}\ \emph {et~al.}(2007)\citenamefont {Young},
  \citenamefont {Urbano}, \citenamefont {Curro}, \citenamefont {Thompson},
  \citenamefont {Sarrao}, \citenamefont {Vorontsov},\ and\ \citenamefont
  {Graf}}]{Young2007}%
  \BibitemOpen
  \bibfield  {author} {\bibinfo {author} {\bibfnamefont {B.~L.}\ \bibnamefont
  {Young}}, \bibinfo {author} {\bibfnamefont {R.~R.}\ \bibnamefont {Urbano}},
  \bibinfo {author} {\bibfnamefont {N.~J.}\ \bibnamefont {Curro}}, \bibinfo
  {author} {\bibfnamefont {J.~D.}\ \bibnamefont {Thompson}}, \bibinfo {author}
  {\bibfnamefont {J.~L.}\ \bibnamefont {Sarrao}}, \bibinfo {author}
  {\bibfnamefont {A.~B.}\ \bibnamefont {Vorontsov}},\ and\ \bibinfo {author}
  {\bibfnamefont {M.~J.}\ \bibnamefont {Graf}},\ }\bibfield  {title} {\bibinfo
  {title} {Microscopic evidence for field-induced magnetism in {CeCoIn$_5$}},\
  }\href {https://doi.org/10.1103/PhysRevLett.98.036402} {\bibfield  {journal}
  {\bibinfo  {journal} {Phys. Rev. Lett.}\ }\textbf {\bibinfo {volume} {98}},\
  \bibinfo {pages} {036402} (\bibinfo {year} {2007})}\BibitemShut {NoStop}%
\end{thebibliography}%

\newpage
\section*{Supplemental Information}

\subsection*{Sample Characteristics}

\begin{figure}[t]
\begin{center}
\includegraphics[width=\linewidth]{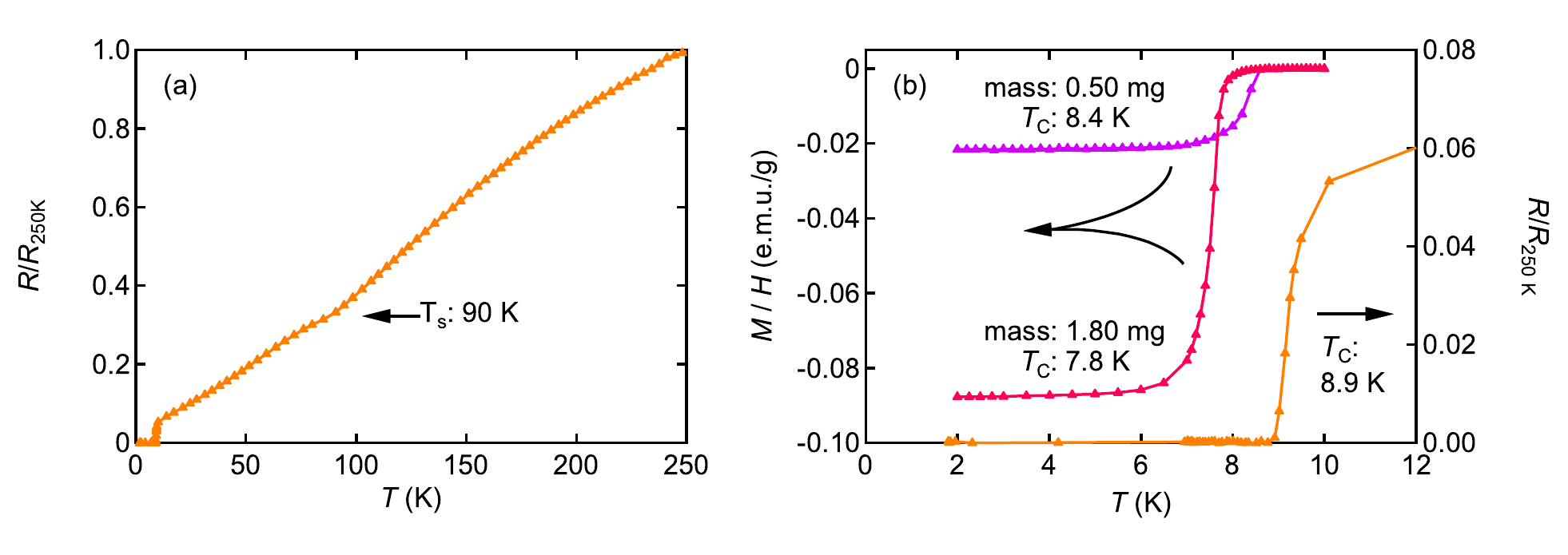}
\caption{(a) Temperature dependent resistivity curve showing the structural transition, $T_s\approx\,90$\,K and a $T_c=8.9$\,K. (b) A zoom in showing the superconducting transition with magnetic susceptibility (purple and red curves) and resistivity (orange curve) data.}
\label{fig:properties}
\end{center}
\end{figure}

Single crystals of FeSe were grown using a vapor transport technique with an angled two-zone furnace \cite{Boehmer2016}. Single crystal samples were characterized via resisitivity and magnetic susceptibility measurements  (Fig.~\ref{fig:properties}). The resistivity data ($R/R_{250 K}$) clearly shows the structural transition $T_{s}\approx\,90$\,K and has a superconducting transition $T_{c}=8.9$\,K. These temperatures and a residual resistivity ratio (RRR$_{250 K}$) value ($18.8$) show that our samples are of similar quality to the ones previously reported. As first noted by B\"ohmer, the offset of resistivity superconducting transition corresponds well with the initial downturn of the magnetic susceptibility. We also note that we found a slight inverse relationship between sample size and transition temperature (Fig.~\ref{fig:properties}(b)). This relationship combined with the slight oxidation of storing the sample in an Ar-glovebox for a year explains the slightly lower superconducting transition of the sample measured in NMR.

\subsection*{Knight shift at Different Power Levels}

\begin{figure}[t]
\begin{center}
\includegraphics[width=\linewidth]{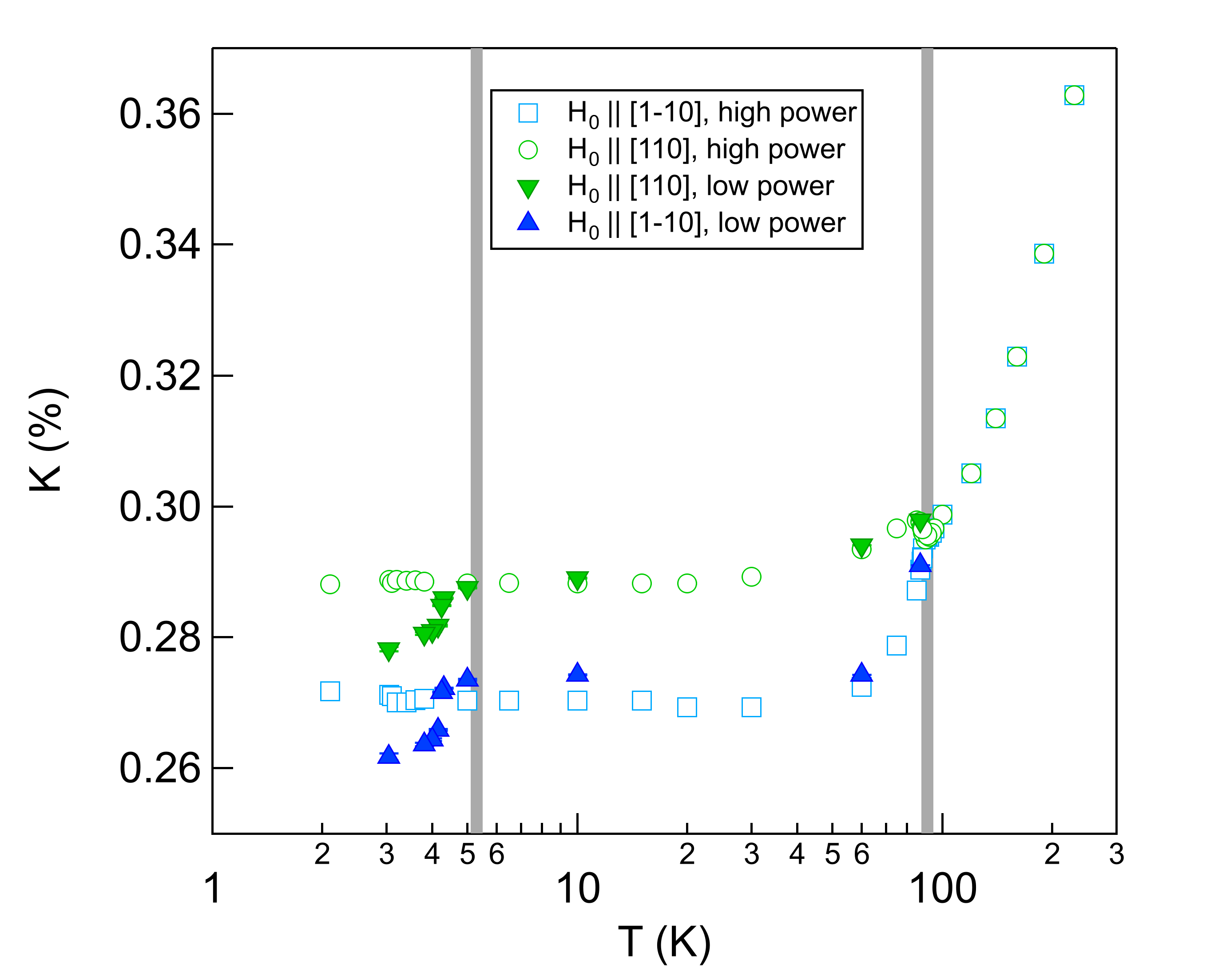}
\caption{(Knight shifts along the $a$ and $b$ directions measured at low and high power. The reduction below $T_c$ is only evident with low power pulses, such that there is no Joule heating of the sample. Differences in the shift above $T_c$ between high and low power spectra are due to slightly worse alignment with [110]/[1$\bar{1}$0] for the low power spectra. }
\label{fig:Kshifts}
\end{center}
\end{figure}

The left panel of Fig. \ref{fig:Kshifts} shows the Knight shift versus temperature for different power levels.  In the first case, the spectra were measured at high power and there is no obvious change below $T_c$ due rf heating effects.  At low power, the shifts are reduced in the superconducting state.

\subsection*{Vortex Lineshape and Knight shift Inhomogeneity}

\begin{figure}[t]
\begin{center}
\includegraphics[width=\linewidth]{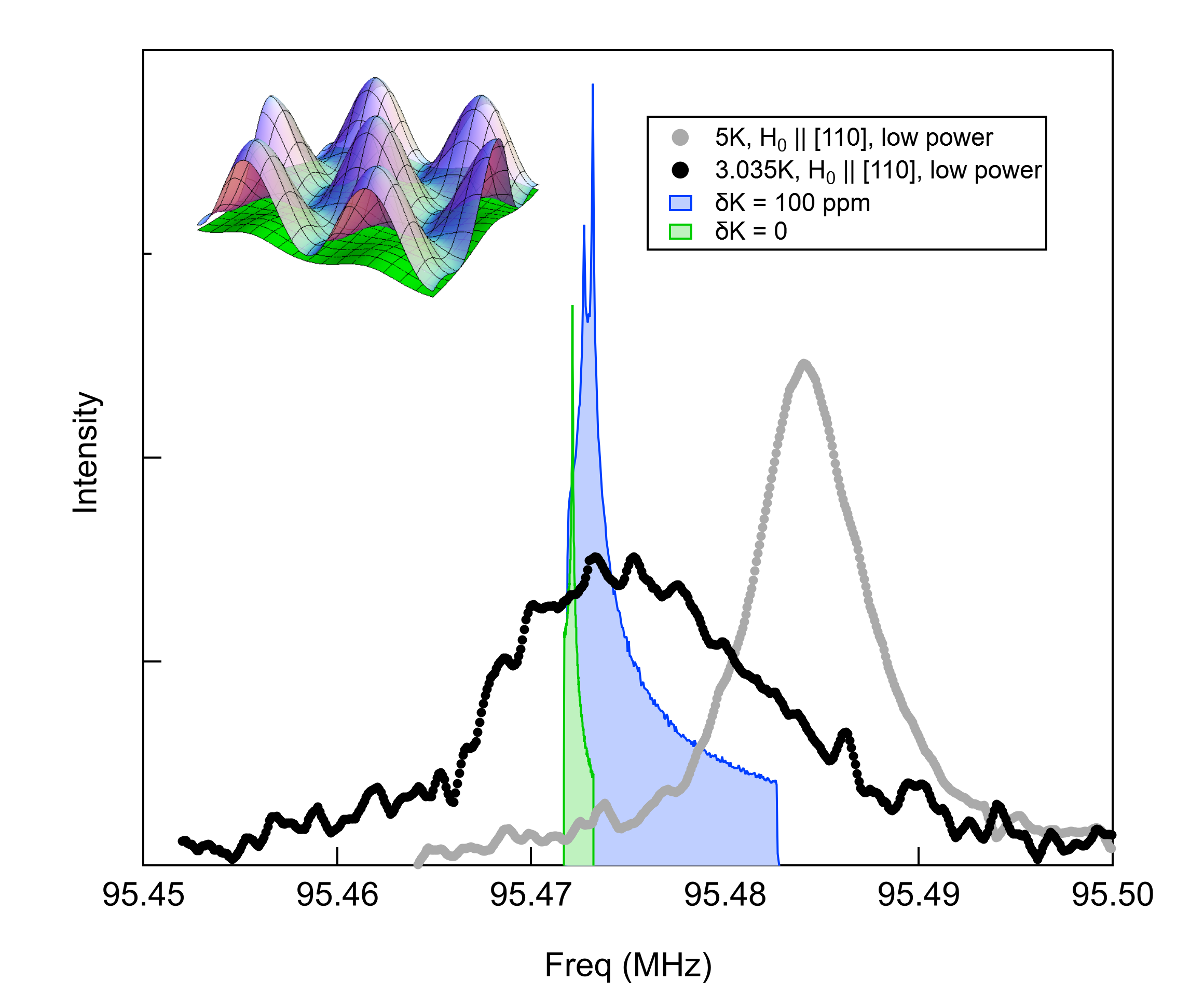}
\caption{$^{77}$Se NMR spectra in the normal and superconducting states at 12 T, measured at low power for field along the [110] direction.  The solid green and blue regions indicate the theoretical spectra computed for a vortex lattice with and without a finite Knight shift, $\delta K$, in the cores, respectively, as discussed above.  The inset shows how the resonance frequency varies as a function of position with and without $\delta K$.  }
\label{fig:VLspectra}
\end{center}
\end{figure}

Figure \ref{fig:VLspectra} compares the spectrum in the normal and superconducting state for  $\mathbf{H}_0~||~ [110]$. In order to model the lineshape in the mixed phase, we use a Monte Carlo approach to compute the histogram, $P(f)$, of local resonance frequencies in a hexagonal vortex lattice, where the frequency is given by $f(\mathbf{r}) = \gamma B(\mathbf{r})(1 + K(\mathbf{r}))$. The local field is given by the London model with a Gaussian cutoff \cite{Maisuradze2009}:
\begin{equation}\label{eqn:London}
  B(\mathbf{r}) = H_0\sum_{\mathbf{G}}\frac{e^{-i\mathbf{G}\cdot\mathbf{r}}}{1 + (\mathbf{\lambda}\cdot\mathbf{G})^2}e^{-|\mathbf{G}|^2\xi^2/2},
\end{equation}
$\mathbf{G}$ are the reciprocal lattice vectors for a hexagonal vortex lattice,  $\mathbf{\lambda} = \left(\lambda_a,\lambda_c\right)$ with $\lambda_a = 446$ nm, $\lambda_c = 1320$ nm, and $\xi = 3.1$ nm \cite{AbdelHafiez2013}.   If we assume that $K(\mathbf{r}) = 0$, then $P(f)$ is given by the green curve in Fig. \ref{fig:VLspectra}.  Although this spectrum does have a high frequency tail, it remains too small to explain the observed inhomogeneity.

To model the spatial dependence of $K(\mathbf{r})$, we assume that it exhibits the periodicity of the vortex lattice with maxima within the cores and vanishing outside, with the expression:
\begin{equation}\label{eqn:shift}
  K(\mathbf{r}) = \delta K\sum_{G}e^{-|\mathbf{G}|^2\xi^2/2}\left(e^{-i\mathbf{G}\cdot\mathbf{r}}-e^{-i\mathbf{G}\cdot\mathbf{r}_0}\right)/K_0,
\end{equation}
where $K_0 = \sum_{\mathbf{G}'}\exp(-|\mathbf{G}'|^2\xi^2/2)\left(1-e^{-i\mathbf{G}'\cdot\mathbf{r}_0}\right)$, $\mathbf{r}_0 = \{a/2,\sqrt{3}a/6\}$ is the location of the field minimum between the vortex cores,  $a$ is the unit cell length for the hexagonal vortex lattice, and $\delta K$ is the shift within the cores. This distribution is illustrated in the inset of Fig. \ref{fig:VLspectra} for $\delta K = 100$ ppm.  In this case, $P(f)$, shown in blue in Fig. \ref{fig:VLspectra}, is broader and  the high frequency tail is extended up to the normal state resonance frequency.

\end{document}